\documentclass[aps,prl,superscriptaddress,twocolumn]{revtex4-2}

\usepackage{graphicx}
\usepackage{amsmath}
\usepackage{amsthm}
\usepackage{amssymb}
\usepackage{hyperref}
\usepackage[utf8]{inputenc}
\usepackage{mathtools}
\usepackage[english]{babel}
\usepackage{bbm}
\hypersetup{colorlinks=true, linkcolor=blue, citecolor=blue, urlcolor=blue}
\usepackage{xcolor}
\usepackage[normalem]{ulem}
\usepackage{braket}

\newcommand{\bdag}{b^{\dag}}

\renewcommand{\P}{\mathcal{P}}

\newcommand{\me}{\mathrm{e}}
\newcommand{\abs}[1]{\lvert #1 \rvert} % absolute value
\DeclareMathOperator{\Tr}{Tr}

\theoremstyle{plain}

\theoremstyle{definition}

\begin{document}

\title{Chirality Dependent Photon Transport and Helical Superradiance}

\author{Jonah S. Peter}
\email{jonahpeter@g.harvard.edu}
\affiliation{Department of Physics, Harvard University, Cambridge, Massachusetts 02138, USA}
\affiliation{Biophysics Program, Harvard University, Boston, Massachusetts 02115, USA}
\author{Stefan Ostermann}
\affiliation{Department of Physics, Harvard University, Cambridge, Massachusetts 02138, USA}
\author{Susanne F. Yelin}
\affiliation{Department of Physics, Harvard University, Cambridge, Massachusetts 02138, USA}

\begin{abstract}
Chirality, or handedness, is a geometrical property denoting a lack of mirror symmetry. Chirality is ubiquitous in nature and is associated with the non-reciprocal interactions observed in complex systems ranging from biomolecules to topological materials. Here, we demonstrate that chiral arrangements of dipole-coupled atoms or molecules can facilitate the unidirectional transport of helical photonic excitations without breaking time-reversal symmetry. We show that such helicity dependent transport stems from an emergent spin-orbit coupling induced by the chiral geometry, which results in nontrivial topological properties. We also examine the effects of collective dissipation and find that many-body coherences lead to helicity dependent photon emission: an effect we call helical superradiance. Our results demonstrate an intimate connection between chirality, topology, and photon helicity that may contribute to molecular photodynamics in nature and could be probed with near-term quantum simulators.
\end{abstract}

\maketitle

%==============================================================================================

Chirality, or handedness, is ubiquitous in nature and can be observed in macromolecular structures like DNA, down to the single particle level as in photons with circular polarization. An object is chiral if it cannot be superimposed on its mirror image by a rotation. This property allows chiral systems to facilitate the unidirectional propagation of photons or charge currents. Platforms that admit such chiral (or related ``helical") modes, including photonic nanostructures \cite{lodahl_chiral_2017, wang_band_2019, segev_topological_2020, ozawa_topological_2019, bliokh_spinorbit_2015} and topological insulators \cite{kane_z_2005, kane_quantum_2005, hasan_colloquium_2010}, are at the forefront of recent innovations in both fundamental physics and applied quantum technologies.

Recently, chiral molecules have emerged as an attractive platform for the development of spintronics devices. The Chiral-Induced Spin Selectivity (CISS) effect~\cite{ray_asymmetric_1999, gohler_spin_2011}---which results in the spin polarization of electrons moving through chiral molecules---has promising applications in magnetless spin memories and spin-based logic gates \cite{naaman_spintronics_2015, michaeli_new_2017, ben_dor_magnetization_2017}, as well as in electrochemistry \cite{naaman_chiral_2020, abendroth_spin_2019}. Additional proposals have suggested that spin selective transport may be involved in biological processes \cite{carmeli_spin_2014, michaeli_electrons_2016} and could have contributed to the emergence of ``biological homochirality" (the observation that DNA and other biomolecules exist almost exclusively with one handedness) during the origin of life \cite{naaman_chiral_2019, ozturk_origins_2022}.

Of course in addition to electrons, photons can also carry spin angular momentum, which is encoded in their two orthogonal polarizations. The coupling of photons to atoms or molecules~\cite{lehmberg_radiation_1970, lehmberg_radiation_1970_2} can result in efficient excitation transport~\cite{olmos_long-range_2013, gutierrez-jauregui_directional_2022, jang_delocalized_2018} and cooperative phenomena including the superradiant and subradiant emission of light~\cite{asenjo-garcia_exponential_2017, reitz_cooperative_2022}. Demonstration of chirality-induced \emph{photon} transport could facilitate an analogous explosion in the development of helicity dependent photonics devices and help unveil the role of chiral light-matter interactions in natural (bio)molecules.
\begin{figure}[htbp]
\centering 
\renewcommand\figurename{FIG.}
\includegraphics[width=\columnwidth]{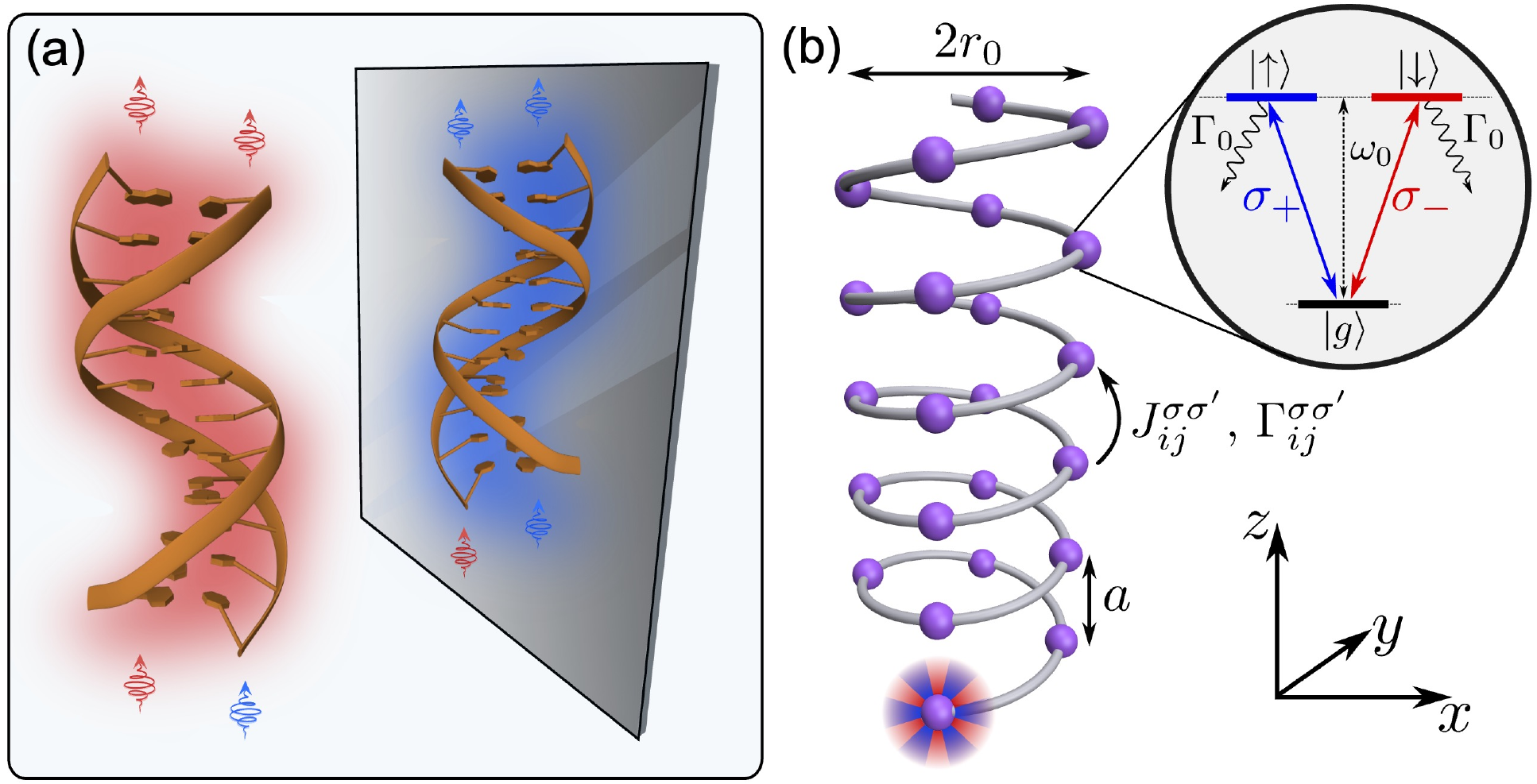}
\caption{(a) Illustration of the main results of this work. Photons with a given helicity (red or blue) are preferentially transported through structures of a given chirality, leading to helical superradiance~\cite{[{Structural image from the RCSB PDB (RCSB.org) of PDB ID 1FZX}]macdonald_solution_2001}. (b) The model used to describe the transport dynamics. Purple spheres denote emitters with a V-type level structure and spontaneous emission rate $\Gamma_0$ (inset). The lattice geometry is a helix with radius $r_0$, pitch $a$, and $\mathcal{N}$ atoms per $2\pi$ turn. Emitters are coupled with coherent ($J_{ij}^{\sigma\sigma'}$) and dissipative ($\Gamma_{ij}^{\sigma\sigma'}$) hopping rates. The multicolored halo denotes initialization in an unpolarized mixed state.}
\label{fig_1}
\end{figure}

In this Letter, we show that helical excitations induced by circularly polarized photons propagate with a helicity dependence through chiral arrangements of dipolar quantum emitters. This phenomenon---which occurs at zero magnetic field and without breaking time-reversal ($\mathcal{T}$) symmetry---results from an emergent spin-orbit coupling (SOC) that is unique to chiral geometries. The dynamics are further enriched in the presence of dissipation where many-body coherences result in helicity dependent photon emission: an effect we call helical superradiance [Fig. \ref{fig_1}(a)]. We show that helicity dependent chiral transport is associated with a non-Abelian gauge field \cite{wilczek_appearance_1984} that results in an excitation band structure with nontrivial topology. Although previous studies have investigated topological properties within photonics systems \cite{segev_topological_2020, ozawa_topological_2019, hafezi_robust_2011, syzranov_spinorbital_2014}, none have explored the connection to geometrical chirality. These effects may contribute to chiral molecular processes in nature and provide a new framework for studying photoexcitation dynamics in chiral molecules using cold atom quantum simulators.

\begin{figure*}[htbp]
\centering 
\renewcommand\figurename{FIG.}
\includegraphics[width=\textwidth]{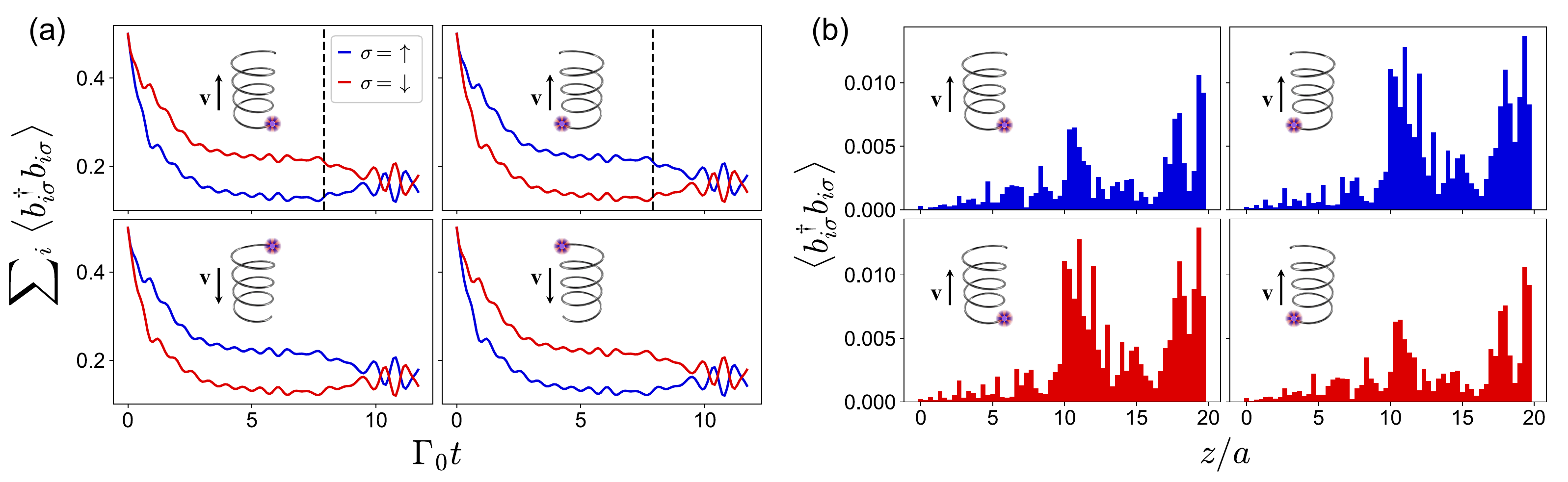}
\caption{(a) Population dynamics for left-handed (left-hand panels) and right-handed (right-hand panels) helices after initialization at either the bottom (top two panels) or top (bottom two panels) of the helix. Blue (red) curves denote the total population in the $\uparrow$ ($\downarrow$) manifold as a function of time. Black arrows denote the propagation direction of the initial wave packet. Dashed lines indicate the time $t = \tau$ at which the wave packet reaches the opposite end of the helix. (b) Snapshots of the individual emitter at the time denoted in panel (a). Blue (red) bars denote the spin $\uparrow$ ($\downarrow$) populations of each emitter along the longitudinal helical axis. Relevant parameters:  $r_0 = 0.05 \lambda_0$, $a = 0.175 \lambda_0$, $\mathcal{N} = 3$, $M = 20$, and $\tau = 7.9/\Gamma_0$.
}
\label{fig_2}
\end{figure*}

\emph{Helicity dependent dynamics.}---Excitation transport between atoms or molecules can be modeled as a collection of quantum emitters interacting with a radiation field \cite{may_charge_2011}. As a minimal model, we focus on a helix as the archetypal chiral geometry---though our results are generalizable to arbitrary chiral setups (see Ref.~\cite{peter_PRA}). The individual sites of a helical molecular aggregate are modeled as V-type quantum emitters, each with two hyperfine transitions excited by left $(\sigma^+)$ and right $(\sigma^-)$ circularly polarized light, respectively, and resonance frequency $\omega_0 = 2\pi c / \lambda_0$ [Fig.~\ref{fig_1}(b)]. Long-range dipole-dipole interactions between emitters at positions $\mathbf{r}_{i}$ and $\mathbf{r}_{j}$ are mediated through real and virtual photon exchanges that couple orbitals $\ket{\sigma_i}$ and $\ket{\sigma'_j}$ with either the same or opposite polarization. After tracing out the field degrees of freedom in the Born and Markov approximations \cite{asenjo-garcia_exponential_2017} and projecting into the single-excitation manifold, this system can be described as a collection of pseudospin-$1/2$ bosons via the Hamiltonian
\begin{equation}\label{eq:hamiltonian}
    H = \sum_{i=1}^N \sum_{\sigma} \omega_0 \bdag_{i\sigma} b_{i\sigma} + \sum^N_{i,j \neq i = 1} \sum_{\sigma,\sigma'} J_{ij}^{\sigma \sigma'} \bdag_{i\sigma} b_{j\sigma'}
\end{equation}
(we set $\hbar \equiv 1$ here and throughout this work). Here, $\bdag_{i\sigma}$ $(b_{i\sigma})$ creates (annihilates) an excitation at site $i$ with spin $\sigma \in \{\uparrow,\downarrow\}$, $N$ is the total number of emitters, and $J_{ij}^{\sigma \sigma'}$ are the spin-dependent hopping rates.

In addition to the coherent interactions described by Eq.~\eqref{eq:hamiltonian}, we also consider collective dissipation that arises through coupling to the electromagnetic vacuum at zero temperature. In the single-excitation subspace, the full open system dynamics result from non-unitary evolution with the non-Hermitian effective Hamiltonian
\begin{equation}\label{eq:H_eff}
    H_{\mathrm{eff}} = H - i \sum_{i,j=1}^N \frac{\Gamma_{ij}^{\sigma \sigma'}}{2} b_{i\sigma}^\dagger b_{j\sigma'}.
\end{equation}
The anti-Hermitian part of Eq.~\eqref{eq:H_eff} describes cooperative decay to the vacuum with rates $\Gamma_{ij}^{\sigma \sigma'}$ and single-emitter spontaneous emission with rate $\Gamma_0 \equiv \Gamma^{\sigma \sigma}_{ii}$. In free space, the couplings are determined by the electromagnetic Green's tensor, $\mathbf{G}(\mathbf{r}_{i}, \mathbf{r}_{j}, \omega_0)$, as
\begin{equation}\label{eq:inter}
    J^{\sigma\sigma'}_{ij} - \frac{i}{2} \Gamma^{\sigma\sigma'}_{ij} = - \frac{3}{2} \lambda_0 \Gamma_0 \hat{\pmb{\varepsilon}}^{\dagger}_{\sigma} \cdot \mathbf{G}(\mathbf{r}_{i}, \mathbf{r}_{j}, \omega_0) \cdot \hat{\pmb{\varepsilon}}_{\sigma'}
\end{equation}
where $\hat{\pmb{\varepsilon}}_{\uparrow \downarrow} = \hat{\mathbf{x}} \pm \hat{\mathbf{y}} / \sqrt{2}$ denote the unit vectors of circular polarization. The time evolution of a general state $\rho(t)$ is then governed by the no-jump quantum master equation, $\dot{\rho} = -i ( H_{\mathrm{eff}} \rho - \rho H^\dag_{\mathrm{eff}} )$.
\begin{figure*}
\centering 
\renewcommand\figurename{FIG.}
\includegraphics[width=\textwidth]{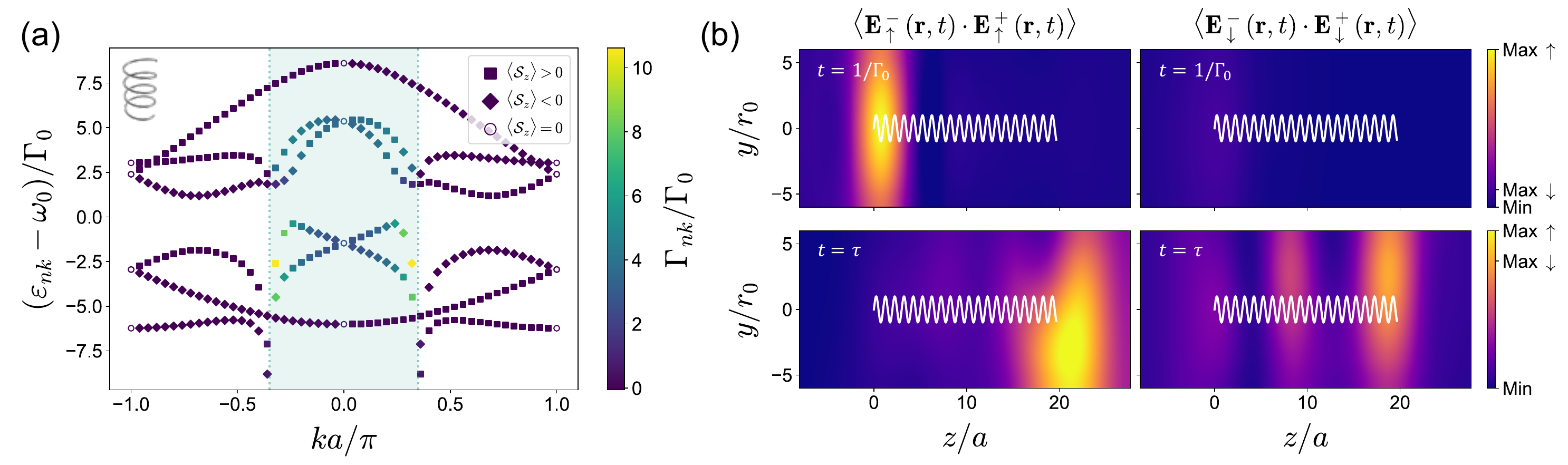}
\caption{(a) Collective excitation band structure for an infinite left-handed helix with the same parameters as in Fig. \ref{fig_2}. Filled squares (diamonds) denote modes that are predominantly spin $\uparrow$ ($\downarrow$). Open circles indicate equal superposition modes. The coloring of each symbol denotes the collective decay rate of each mode, $\Gamma_{nk}$. Dotted lines indicate the light cone at $k=k_0$. (b) Snapshots of the emitted light field intensity at $t = 1/\Gamma_0$ (top two panels) and $t = \tau$ (bottom two panels). The left (right) two panels show the contribution from spin $\uparrow$ ($\downarrow$) photons to the field intensity. Brighter colors demonstrate helical superradiance that is enhanced at either end of the helix. Field intensities are shown in the $y$-$z$ plane at a distance $x = 10 r_0$ from the helical axis so as to not include contributions from the evanescent fields of the emitters. A projection of the helix geometry onto the $y$-$z$ plane is shown in white. Color bars indicate the relative field intensity of each polarization.}
\label{fig_3}
\end{figure*}

To illustrate the dynamics, we consider left- (denoted by $\xi=1)$ and right- ($\xi=-1$) handed helices oriented along the $z$-axis with radius $r_0$, pitch $a$, $\mathcal{N}$ emitters per $2\pi$ turn, and $M = N/\mathcal{N}$ turns total. The initial state is chosen as $\rho(0) = \frac{1}{2}\left(\ket{\uparrow_i} \bra{\uparrow_i} + \ket{\downarrow_i} \bra{\downarrow_i}\right)$, corresponding to an unpolarized statistical mixture of up and down spins localized to a single emitter at $\mathbf{r}_i$. Fig. \ref{fig_2}(a) shows the total population summed across all emitters in each spin manifold, $\sum_i \braket{b^{\dag}_{i\sigma} b_{i\sigma}}$, as a function of time. The top (bottom) two panels show the dynamics when the starting emitter is located at the bottom (top) of the helix. The spin of each state is given by the expectation value $\braket{\mathcal{S}_z}$ where $\mathcal{S}_z = \openone_N \otimes \sigma_z = \sum_i (\bdag_{i\uparrow} b_{i\uparrow} - \bdag_{i\downarrow} b_{i\downarrow})$, $\openone_N$ is the $N \times N$ identity matrix, and $\sigma_z$ is the third Pauli operator. As the initial spin wave propagates through the helix with group velocity $\mathbf{v} = v \hat{\mathbf{z}}$ (black arrows), it acquires a helicity, $\eta = \braket{\mathcal{S}_z} v / \abs{\braket{\mathcal{S}_z} v} = \pm 1$. For the left-handed chirality [left-hand panels of Fig. \ref{fig_2}(a)], the positive helicity states [i.e., spin $\uparrow$ ($\downarrow$) propagating upwards (downwards)] experience enhanced dissipation. By contrast for the right-handed chirality [right-hand panels of Fig. \ref{fig_2}(a)], enhanced dissipation occurs with the negative helicity states [spin $\downarrow$ ($\uparrow$) propagating upwards (downwards)]. In both cases, the product $\chi \equiv \xi \eta  = -1$ describes configurations that are more robust to population loss. Fig.~\ref{fig_2}(b) shows a snapshot of the individual emitter populations when the initial upward-moving wave packet reaches the top end of the helix [time $t = \tau$, denoted by dashed lines in Fig.~\ref{fig_2}(a)]. For the left-handed helix, spin $\downarrow$ excitations exhibit longer lifetimes and are more efficiently transported to the opposite end. The same initial condition results in the preferential transport of spin $\uparrow$ excitations for the right-handed helix. For the more general case when the initial state is an \emph{unequal} statistical mixture of spins, the same transport dynamics result in the predominant excitation of only one chirality (Supplemental Fig. S1).

\emph{Emergent SOC and helical superradiance.}---The transport dynamics described above can be understood by examining the band structure of the collective helical modes [Fig.~\ref{fig_3}(a)]. Eq.~\eqref{eq:H_eff} can be expressed in momentum space by performing the discrete Fourier transform $b_{i\sigma} = (1/\sqrt{M}) \sum_\mathbf{k} \exp{(i\mathbf{k}\cdot\mathbf{r}_i)} b_{k\mu\sigma}$, where $\mu$ is the sublattice index denoting the $M$ emitters along one sublattice and $\mathbf{k} = k \hat{\mathbf{z}}$ is the lattice quasimomentum \cite{peter_PRA}. The eigenstates of the resultant $k$-space Hamiltonian are collective Bloch modes of the form $\ket{\psi_{n k}} = \me^{i k z} \ket{u_{nk}}$ with complex eigenvalues $\tilde{\varepsilon}_{nk} = \varepsilon_{nk} + i \Gamma_{nk}$ and band index $n$. Here, $\varepsilon_{nk}$ is the energy of each Bloch mode, and $\Gamma_{nk}$ is the corresponding vacuum decay rate, as indicated by the color coding of symbols in Fig.~\ref{fig_3}(a). Eq.~\eqref{eq:H_eff} breaks spin rotation symmetry through the term $\propto b_{i,\uparrow}^\dagger b_{j,\downarrow}$ but is invariant under the combined operation of spatial inversion and spin-flip. Denoting the usual parity operator as $\P$, this ``anti-inversion" symmetry can be written as $\bar{\P} = \P \otimes \sigma_x$ and results in antisymmetric spin textures for the Bloch bands. $\bar{\P}$ transforms $(k, \mathcal{S}_z) \to (-k, -\mathcal{S}_z)$ and therefore requires modes with quasimomentum $\pm k$ to have equal energy but opposite spin and group velocity $v = d\varepsilon_k/dk$. Besides the $\bar{\P}$-invariant points at $k = 0, \pm \pi/a$ (at which $\braket{\mathcal{S}_z} = 0$), broken spin rotation symmetry allows each mode to experience spin mixing. The spin of each Bloch mode is allowed to be nonzero for arbitrary $k \neq 0, \pm \pi/a$ when the geometry is chiral. This property allows modes with finite dispersion ($v \neq 0$) to experience SOC and results in more efficient transport of spin excitations in the $\chi = -1$ configuration. As we show in Ref.~\cite{peter_PRA}, the nontrivial spin dynamics result from the broken mirror symmetry of the system and are a general feature of arbitrary chiral geometries. 

\begin{figure*}[t]
\centering 
\renewcommand\figurename{FIG.}
\includegraphics[width=\textwidth]{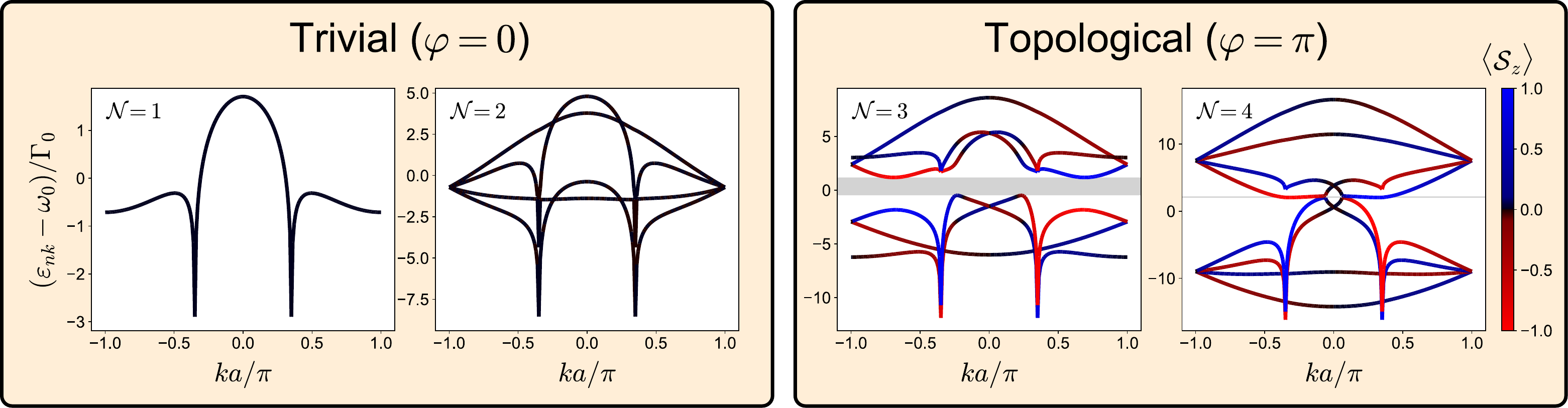}
\caption{Demonstration of nontrivial topology for chiral geometries. Each plot shows the collective band structure for a left-handed helix constructed with $\mathcal{N}$ emitters per $2\pi$ turn. The radius and pitch are the same as in Figs. \ref{fig_2} and \ref{fig_3}. A topological phase transition occurs at $\mathcal{N} \geq 3$, characterized by the opening of an energy band gap (gray shaded regions) and a nontrivial Zak phase, $\varphi$. Colors show the spin character $\braket{\mathcal{S}_z}$ of each mode and demonstrate the emergence of a finite SOC in the topologically nontrivial regime.}
\label{fig_4}
\end{figure*}

The antisymmetric spin textures and associated helicity dependent chiral transport are protected by $\bar{\P}$ symmetry irrespective of the behavior under $\mathcal{T}$ (Supplemental Fig. S2). Nevertheless, inclusion of the anti-Hermitian term in Eq.~\eqref{eq:H_eff} enhances this dynamical effect due to collective dissipation at subwavelength scales. In this regime, cooperative resonances can enhance or suppress the decay rate of a given mode away from the bare emitter decay rate, $\Gamma_0$, depending on whether that mode lies inside or outside the light cone [dotted lines in Fig. \ref{fig_3}(a)]. Bright modes (those inside the light cone) are superradiant with $\Gamma_{nk} > \Gamma_0$, whereas dark modes (outside the light cone) experience a momentum mismatch with the available vacuum modes and are subradiant with $\Gamma_{nk} \approx 0$. Eq.~\eqref{eq:H_eff} breaks $\mathcal{T}$ symmetry but does not alter the symmetry properties of the spin bands under $\bar{\P}$. Moreover, the decay rates of the Bloch modes may be interpreted as imaginary energies resulting from evolution with $H_{\mathrm{eff}}$. Therefore, as for the real energies, symmetry under $\bar{\P}$ requires that $\Gamma_{nk}$ be equal for modes with $\pm k$. Because these modes necessarily have opposite spin and group velocity, they also have identical helicity. As such, the two maximally superradiant modes always enhance the population loss for the same photon helicity. Because mirror reflection transforms $\xi \to - \xi$ and $\eta \to - \eta$, the superradiant helicity is reversed between the left- and right-handed chiralities and amplifies the chiral transport. This dynamical effect results in the preferential population of a given spin manifold determined by the propagation direction and chirality of the geometry.

Fig.~\ref{fig_3}(b) demonstrates the helical superradiance explicitly. For each polarization, the positive frequency electric dipole field at each point $\mathbf{r}$ is given by 
\begin{equation}
\mathbf{E}^+_{\sigma}(\mathbf{r}) = \sqrt{\frac{6\pi^2 \Gamma_0}{\lambda_0 \epsilon_0}} \sum_{j=1}^{N} \mathbf{G}(\mathbf{r} - \mathbf{r}_{j}, \omega_0) \cdot \hat{\pmb{\varepsilon}}_{j\sigma} \ket{g} \bra{\sigma_j},
\end{equation}
where $\ket{g}$ is the collective ground state and $\epsilon_0$ is the vacuum permittivity. The left (right) panels show the intensity of the emitted field, $\braket{\mathbf{E}^-_{\sigma}(\mathbf{r}, t) \cdot \mathbf{E}^+_{\sigma}(\mathbf{r}, t)}$, radiated by $\uparrow$ ($\downarrow$) excitations. The top two panels show the initial superradiant burst at $t = 1/\Gamma_0$ after initialization with an unpolarized mixed state at the bottom end of the helix [see also top left panel of Fig.~\ref{fig_2}(a)]. The intense emission of $\uparrow$-polarized photons results from dynamically generated overlaps between the evolving state, $\rho(t)$, and the positive helicity superradiant modes of Fig.~\ref{fig_3}(a). At intermediate times, the wave packet enters subradiant guided modes \cite{asenjo-garcia_exponential_2017} and radiation into the vacuum via the bulk emitters is suppressed. Strong superradiant emission resumes at $t = \tau$ out of the opposite end of the helix [bottom two panels of Fig \ref{fig_3}(b)].

\emph{Topological properties.}---The SOC identified above implicates a nontrivial topology for the energy bands of chiral geometries \cite{kane_topological_2013, halperin_quantized_1982, kane_quantum_2005, atala_direct_2013}. Conservation of angular momentum requires that the two-body spin-flip process in Eq.~\eqref{eq:inter} picks up a complex phase $J_{ij}^{\uparrow \downarrow} \propto \exp{\{-2i(\phi_{i} + \phi_j)\}}$ related to the azimuthal positions of the two emitters in order to compensate for the changing photon polarization. This phase dependence, which stems from the electromagnetic Green's tensor, gives rise to an emergent gauge field that results in a nontrivial topology. In 1D, the topological properties of the band structure are characterized by the Zak phase~\cite{zak_berrys_1989},
\begin{equation}
    \varphi = \oint_{\mathcal{C}} \Tr{[\mathbf{A}(k)]} dk,
    \label{eq:gauge_field}
\end{equation}
where $A_{mn}(k) = i \braket{u_{mk} | \partial_k u_{nk}}$ is the non-Abelian Berry connection \cite{wilczek_appearance_1984} and $\mathcal{C}$ defines a closed loop in reciprocal space around the Brillouin zone torus. The Zak phase for a parity (anti)symmetric unit cell is defined modulo $2\pi$ and can be either $0$ (trivial) or $\pi$ (nontrivial).

Fig.~\ref{fig_4} demonstrates a transition to topologically nontrivial phases as a function of $\mathcal{N}$. When $\mathcal{N}=1$, the lattice is a simple 1D chain of emitters with lattice spacing $a$. In this case, the two spin manifolds are uncoupled with $J_{ij}^{\uparrow \downarrow} = 0$, and the band structure consists of two degenerate copies of the band structure for a chain of transversely polarized two-level emitters~\cite{asenjo-garcia_exponential_2017}. Spin rotation symmetry requires that the two degenerate states at each $k$ have $\braket{\mathcal{S}_z} = \pm 1$, which averages to zero at every point in the Brillouin zone. For $\mathcal{N}=2$, the geometry is that of a staggered chain with longitudinal separation $a$ and transverse separation $2r_0$. Here, uniform spin mixing occurs at each mode through $J_{ij}^{\uparrow \downarrow} \neq 0$, and the band structure resembles that of two 1D chains separated by a finite interaction energy. The presence of reflection symmetry for $\mathcal{N} = 1, 2$ prohibits the emergence of a finite SOC \cite{peter_PRA}. The minimal chiral geometry is realized for $\mathcal{N} = 3$, resulting in a finite energy gap and SO coupled bands. The Zak phases calculated on either side of the energy gap are nontrivial, indicating the transition to a topologically distinct phase. This nontrivial topology persists for $\mathcal{N} \geq 3$ as long as the geometry remains chiral~\cite{peter_PRA}.

These topological considerations might also be relevant to the spin selective transport of helical electrons in chiral molecules~\cite{liu_chirality-driven_2021}. Presently, there is no consensus explanation for the extremely large spin polarizations produced by the CISS effect. However, a chirality-induced gauge field like that in Eq.~\eqref{eq:gauge_field} has not been considered in any theoretical analysis of electron spin polarization in chiral molecules thus far. Our results further suggest that dissipation is crucial for achieving strong helicity dependent transport. The preferential scattering of excitations with the ``wrong" helicity (i.e., helical superradiance) greatly enhances this effect as compared to the fully Hermitian case (Supplemental Fig. S2). Accounting for topologically protected spin polarizations and open system dynamics might resolve the discrepancies between theoretical and experimental models of CISS.

\emph{Outlook.}---Here we have demonstrated helicity dependent photonic excitation transport where the internal emitter degrees of freedom are coupled to an underlying chiral geometry. This phenomenon has a complete description within the electric dipole approximation, placing it within the new class of nonmagnetic chiral interactions exhibiting very strong optical responses \cite{ayuso_ultrafast_2022, ordonez_generalized_2018}. Precise control over the transport and emission of photons is a fundamental goal of quantum information science and could contribute to the development of new quantum technologies. Our findings represent an exciting new avenue for cold atom quantum simulators towards studying chiral light-matter interactions in a well-controlled setting. The exploitation of photon helicity as an additional degree of freedom might also allow for new types of chiral-selective chemistry driven by circularly polarized light, without reliance on higher order multipole moments. 

These phenomena may also play a role in fundamental processes in nature. The well-documented importance of photoexcitation dynamics to the stability and reactivity of (pre)biotic molecules (e.g., RNA, DNA) \cite{nielsen_probing_2012, nielsen_electronic_2013, buchvarov_electronic_2007} evokes the question of whether the effects presented here might have contributed to the emergence of biological homochirality. Although the mechanisms responsible for the homochirality of life are unknown, recent studies have suggested that chiral selective processes during prebiotic synthesis could have seeded an initial imbalance that was amplified by subsequent chemical reactions \cite{blackmond_origin_2010, hein_origin_2012}. In particular, a bias in the net flux of circularly polarized photons---as has been observed in nearby star forming regions \cite{bailey_circular_1998, fukue_extended_2010, kwon_near-infrared_2013}---has long been suggested as a potential symmetry breaking agent \cite{hadidi_electron_2018, garcia_astrophysical_2019, jorissen_asymmetric_2002}. However, previous attempts to translate this bias into homochiral chemistry using traditional (magnetic) chiroptical phenomena typically result in asymmetries of only a few percent~\cite{balavoine_preparation_1974, flores_asymmetric_1977, meierhenrich_asymmetric_2005, meinert_photonenergy-controlled_2014}. The helical superradiance identified in this work provides a new mechanism for chiral amplification that does not rely on weak magnetic interactions and could potentially lead to a much larger chiral bias.
\linebreak\par
The authors thank Mikhail D. Lukin and Jonathan Simon for discussions on quantum simulators, Dimitar D. Sasselov and Rafal Szabla for discussions on chiral molecules, and Ceren B. Dag for discussions on topological band theory. S.O. is supported by a postdoctoral fellowship of the Max Planck-Harvard Research Center for Quantum Optics. All authors acknowledge funding from the National Science Foundation (NSF) via the Center for Ultracold Atoms (CUA) Physics Frontiers Centers (PFC) program and via PHY-2207972, as well as from the Air Force Office of Scientific Research (AFOSR).

\bibliography{references}

\begin{thebibliography}{54}%
\makeatletter
\providecommand \@ifxundefined [1]{%
 \@ifx{#1\undefined}
}%
\providecommand \@ifnum [1]{%
 \ifnum #1\expandafter \@firstoftwo
 \else \expandafter \@secondoftwo
 \fi
}%
\providecommand \@ifx [1]{%
 \ifx #1\expandafter \@firstoftwo
 \else \expandafter \@secondoftwo
 \fi
}%
\providecommand \natexlab [1]{#1}%
\providecommand \enquote  [1]{``#1''}%
\providecommand \bibnamefont  [1]{#1}%
\providecommand \bibfnamefont [1]{#1}%
\providecommand \citenamefont [1]{#1}%
\providecommand \href@noop [0]{\@secondoftwo}%
\providecommand \href [0]{\begingroup \@sanitize@url \@href}%
\providecommand \@href[1]{\@@startlink{#1}\@@href}%
\providecommand \@@href[1]{\endgroup#1\@@endlink}%
\providecommand \@sanitize@url [0]{\catcode `\\12\catcode `\$12\catcode
  `\&12\catcode `\#12\catcode `\^12\catcode `\_12\catcode `\%12\relax}%
\providecommand \@@startlink[1]{}%
\providecommand \@@endlink[0]{}%
\providecommand \url  [0]{\begingroup\@sanitize@url \@url }%
\providecommand \@url [1]{\endgroup\@href {#1}{\urlprefix }}%
\providecommand \urlprefix  [0]{URL }%
\providecommand \Eprint [0]{\href }%
\providecommand \doibase [0]{https://doi.org/}%
\providecommand \selectlanguage [0]{\@gobble}%
\providecommand \bibinfo  [0]{\@secondoftwo}%
\providecommand \bibfield  [0]{\@secondoftwo}%
\providecommand \translation [1]{[#1]}%
\providecommand \BibitemOpen [0]{}%
\providecommand \bibitemStop [0]{}%
\providecommand \bibitemNoStop [0]{.\EOS\space}%
\providecommand \EOS [0]{\spacefactor3000\relax}%
\providecommand \BibitemShut  [1]{\csname bibitem#1\endcsname}%
\let\auto@bib@innerbib\@empty
%</preamble>
\bibitem [{\citenamefont {Lodahl}\ \emph {et~al.}(2017)\citenamefont {Lodahl},
  \citenamefont {Mahmoodian}, \citenamefont {Stobbe}, \citenamefont
  {Rauschenbeutel}, \citenamefont {Schneeweiss}, \citenamefont {Volz},
  \citenamefont {Pichler},\ and\ \citenamefont {Zoller}}]{lodahl_chiral_2017}%
  \BibitemOpen
  \bibfield  {author} {\bibinfo {author} {\bibfnamefont {P.}~\bibnamefont
  {Lodahl}}, \bibinfo {author} {\bibfnamefont {S.}~\bibnamefont {Mahmoodian}},
  \bibinfo {author} {\bibfnamefont {S.}~\bibnamefont {Stobbe}}, \bibinfo
  {author} {\bibfnamefont {A.}~\bibnamefont {Rauschenbeutel}}, \bibinfo
  {author} {\bibfnamefont {P.}~\bibnamefont {Schneeweiss}}, \bibinfo {author}
  {\bibfnamefont {J.}~\bibnamefont {Volz}}, \bibinfo {author} {\bibfnamefont
  {H.}~\bibnamefont {Pichler}},\ and\ \bibinfo {author} {\bibfnamefont
  {P.}~\bibnamefont {Zoller}},\ }\href {https://doi.org/10.1038/nature21037}
  {\bibfield  {journal} {\bibinfo  {journal} {Nature}\ }\textbf {\bibinfo
  {volume} {541}},\ \bibinfo {pages} {473} (\bibinfo {year}
  {2017})}\BibitemShut {NoStop}%
\bibitem [{\citenamefont {Wang}\ \emph {et~al.}(2019)\citenamefont {Wang},
  \citenamefont {Guo},\ and\ \citenamefont {Jiang}}]{wang_band_2019}%
  \BibitemOpen
  \bibfield  {author} {\bibinfo {author} {\bibfnamefont {H.-X.}\ \bibnamefont
  {Wang}}, \bibinfo {author} {\bibfnamefont {G.-Y.}\ \bibnamefont {Guo}},\ and\
  \bibinfo {author} {\bibfnamefont {J.-H.}\ \bibnamefont {Jiang}},\ }\href
  {https://doi.org/10.1088/1367-2630/ab3f71} {\bibfield  {journal} {\bibinfo
  {journal} {New Journal of Physics}\ }\textbf {\bibinfo {volume} {21}},\
  \bibinfo {pages} {093029} (\bibinfo {year} {2019})}\BibitemShut {NoStop}%
\bibitem [{\citenamefont {Segev}\ and\ \citenamefont
  {Bandres}(2020)}]{segev_topological_2020}%
  \BibitemOpen
  \bibfield  {author} {\bibinfo {author} {\bibfnamefont {M.}~\bibnamefont
  {Segev}}\ and\ \bibinfo {author} {\bibfnamefont {M.~A.}\ \bibnamefont
  {Bandres}},\ }\href {https://doi.org/10.1515/nanoph-2020-0441} {\bibfield
  {journal} {\bibinfo  {journal} {Nanophotonics}\ }\textbf {\bibinfo {volume}
  {10}},\ \bibinfo {pages} {425} (\bibinfo {year} {2020})}\BibitemShut
  {NoStop}%
\bibitem [{\citenamefont {Ozawa}\ \emph {et~al.}(2019)\citenamefont {Ozawa},
  \citenamefont {Price}, \citenamefont {Amo}, \citenamefont {Goldman},
  \citenamefont {Hafezi}, \citenamefont {Lu}, \citenamefont {Rechtsman},
  \citenamefont {Schuster}, \citenamefont {Simon}, \citenamefont {Zilberberg},\
  and\ \citenamefont {Carusotto}}]{ozawa_topological_2019}%
  \BibitemOpen
  \bibfield  {author} {\bibinfo {author} {\bibfnamefont {T.}~\bibnamefont
  {Ozawa}}, \bibinfo {author} {\bibfnamefont {H.~M.}\ \bibnamefont {Price}},
  \bibinfo {author} {\bibfnamefont {A.}~\bibnamefont {Amo}}, \bibinfo {author}
  {\bibfnamefont {N.}~\bibnamefont {Goldman}}, \bibinfo {author} {\bibfnamefont
  {M.}~\bibnamefont {Hafezi}}, \bibinfo {author} {\bibfnamefont
  {L.}~\bibnamefont {Lu}}, \bibinfo {author} {\bibfnamefont {M.~C.}\
  \bibnamefont {Rechtsman}}, \bibinfo {author} {\bibfnamefont {D.}~\bibnamefont
  {Schuster}}, \bibinfo {author} {\bibfnamefont {J.}~\bibnamefont {Simon}},
  \bibinfo {author} {\bibfnamefont {O.}~\bibnamefont {Zilberberg}},\ and\
  \bibinfo {author} {\bibfnamefont {I.}~\bibnamefont {Carusotto}},\ }\href
  {https://doi.org/10.1103/RevModPhys.91.015006} {\bibfield  {journal}
  {\bibinfo  {journal} {Reviews of Modern Physics}\ }\textbf {\bibinfo {volume}
  {91}},\ \bibinfo {pages} {015006} (\bibinfo {year} {2019})}\BibitemShut
  {NoStop}%
\bibitem [{\citenamefont {Bliokh}\ \emph {et~al.}(2015)\citenamefont {Bliokh},
  \citenamefont {Rodríguez-Fortuño}, \citenamefont {Nori},\ and\
  \citenamefont {Zayats}}]{bliokh_spinorbit_2015}%
  \BibitemOpen
  \bibfield  {author} {\bibinfo {author} {\bibfnamefont {K.~Y.}\ \bibnamefont
  {Bliokh}}, \bibinfo {author} {\bibfnamefont {F.~J.}\ \bibnamefont
  {Rodríguez-Fortuño}}, \bibinfo {author} {\bibfnamefont {F.}~\bibnamefont
  {Nori}},\ and\ \bibinfo {author} {\bibfnamefont {A.~V.}\ \bibnamefont
  {Zayats}},\ }\href {https://doi.org/10.1038/nphoton.2015.201} {\bibfield
  {journal} {\bibinfo  {journal} {Nature Photonics}\ }\textbf {\bibinfo
  {volume} {9}},\ \bibinfo {pages} {796} (\bibinfo {year} {2015})}\BibitemShut
  {NoStop}%
\bibitem [{\citenamefont {Kane}\ and\ \citenamefont
  {Mele}(2005{\natexlab{a}})}]{kane_z_2005}%
  \BibitemOpen
  \bibfield  {author} {\bibinfo {author} {\bibfnamefont {C.~L.}\ \bibnamefont
  {Kane}}\ and\ \bibinfo {author} {\bibfnamefont {E.~J.}\ \bibnamefont
  {Mele}},\ }\href {https://doi.org/10.1103/PhysRevLett.95.146802} {\bibfield
  {journal} {\bibinfo  {journal} {Physical Review Letters}\ }\textbf {\bibinfo
  {volume} {95}},\ \bibinfo {pages} {146802} (\bibinfo {year}
  {2005}{\natexlab{a}})}\BibitemShut {NoStop}%
\bibitem [{\citenamefont {Kane}\ and\ \citenamefont
  {Mele}(2005{\natexlab{b}})}]{kane_quantum_2005}%
  \BibitemOpen
  \bibfield  {author} {\bibinfo {author} {\bibfnamefont {C.~L.}\ \bibnamefont
  {Kane}}\ and\ \bibinfo {author} {\bibfnamefont {E.~J.}\ \bibnamefont
  {Mele}},\ }\href {https://doi.org/10.1103/PhysRevLett.95.226801} {\bibfield
  {journal} {\bibinfo  {journal} {Physical Review Letters}\ }\textbf {\bibinfo
  {volume} {95}},\ \bibinfo {pages} {226801} (\bibinfo {year}
  {2005}{\natexlab{b}})}\BibitemShut {NoStop}%
\bibitem [{\citenamefont {Hasan}\ and\ \citenamefont
  {Kane}(2010)}]{hasan_colloquium_2010}%
  \BibitemOpen
  \bibfield  {author} {\bibinfo {author} {\bibfnamefont {M.~Z.}\ \bibnamefont
  {Hasan}}\ and\ \bibinfo {author} {\bibfnamefont {C.~L.}\ \bibnamefont
  {Kane}},\ }\href {https://doi.org/10.1103/RevModPhys.82.3045} {\bibfield
  {journal} {\bibinfo  {journal} {Reviews of Modern Physics}\ }\textbf
  {\bibinfo {volume} {82}},\ \bibinfo {pages} {3045} (\bibinfo {year}
  {2010})}\BibitemShut {NoStop}%
\bibitem [{\citenamefont {Ray}\ \emph {et~al.}(1999)\citenamefont {Ray},
  \citenamefont {Ananthavel}, \citenamefont {Waldeck},\ and\ \citenamefont
  {Naaman}}]{ray_asymmetric_1999}%
  \BibitemOpen
  \bibfield  {author} {\bibinfo {author} {\bibfnamefont {K.}~\bibnamefont
  {Ray}}, \bibinfo {author} {\bibfnamefont {S.~P.}\ \bibnamefont {Ananthavel}},
  \bibinfo {author} {\bibfnamefont {D.~H.}\ \bibnamefont {Waldeck}},\ and\
  \bibinfo {author} {\bibfnamefont {R.}~\bibnamefont {Naaman}},\ }\href
  {https://doi.org/10.1126/science.283.5403.814} {\bibfield  {journal}
  {\bibinfo  {journal} {Science}\ }\textbf {\bibinfo {volume} {283}},\ \bibinfo
  {pages} {814} (\bibinfo {year} {1999})}\BibitemShut {NoStop}%
\bibitem [{\citenamefont {Göhler}\ \emph {et~al.}(2011)\citenamefont
  {Göhler}, \citenamefont {Hamelbeck}, \citenamefont {Markus}, \citenamefont
  {Kettner}, \citenamefont {Hanne}, \citenamefont {Vager}, \citenamefont
  {Naaman},\ and\ \citenamefont {Zacharias}}]{gohler_spin_2011}%
  \BibitemOpen
  \bibfield  {author} {\bibinfo {author} {\bibfnamefont {B.}~\bibnamefont
  {Göhler}}, \bibinfo {author} {\bibfnamefont {V.}~\bibnamefont {Hamelbeck}},
  \bibinfo {author} {\bibfnamefont {T.~Z.}\ \bibnamefont {Markus}}, \bibinfo
  {author} {\bibfnamefont {M.}~\bibnamefont {Kettner}}, \bibinfo {author}
  {\bibfnamefont {G.~F.}\ \bibnamefont {Hanne}}, \bibinfo {author}
  {\bibfnamefont {Z.}~\bibnamefont {Vager}}, \bibinfo {author} {\bibfnamefont
  {R.}~\bibnamefont {Naaman}},\ and\ \bibinfo {author} {\bibfnamefont
  {H.}~\bibnamefont {Zacharias}},\ }\href
  {https://doi.org/10.1126/science.1199339} {\bibfield  {journal} {\bibinfo
  {journal} {Science}\ }\textbf {\bibinfo {volume} {331}},\ \bibinfo {pages}
  {894} (\bibinfo {year} {2011})}\BibitemShut {NoStop}%
\bibitem [{\citenamefont {Naaman}\ and\ \citenamefont
  {Waldeck}(2015)}]{naaman_spintronics_2015}%
  \BibitemOpen
  \bibfield  {author} {\bibinfo {author} {\bibfnamefont {R.}~\bibnamefont
  {Naaman}}\ and\ \bibinfo {author} {\bibfnamefont {D.~H.}\ \bibnamefont
  {Waldeck}},\ }\href {https://doi.org/10.1146/annurev-physchem-040214-121554}
  {\bibfield  {journal} {\bibinfo  {journal} {Annual Review of Physical
  Chemistry}\ }\textbf {\bibinfo {volume} {66}},\ \bibinfo {pages} {263}
  (\bibinfo {year} {2015})}\BibitemShut {NoStop}%
\bibitem [{\citenamefont {Michaeli}\ \emph {et~al.}(2017)\citenamefont
  {Michaeli}, \citenamefont {Varade}, \citenamefont {Naaman},\ and\
  \citenamefont {Waldeck}}]{michaeli_new_2017}%
  \BibitemOpen
  \bibfield  {author} {\bibinfo {author} {\bibfnamefont {K.}~\bibnamefont
  {Michaeli}}, \bibinfo {author} {\bibfnamefont {V.}~\bibnamefont {Varade}},
  \bibinfo {author} {\bibfnamefont {R.}~\bibnamefont {Naaman}},\ and\ \bibinfo
  {author} {\bibfnamefont {D.~H.}\ \bibnamefont {Waldeck}},\ }\href
  {https://doi.org/10.1088/1361-648X/aa54a4} {\bibfield  {journal} {\bibinfo
  {journal} {Journal of Physics: Condensed Matter}\ }\textbf {\bibinfo {volume}
  {29}},\ \bibinfo {pages} {103002} (\bibinfo {year} {2017})}\BibitemShut
  {NoStop}%
\bibitem [{\citenamefont {Ben~Dor}\ \emph {et~al.}(2017)\citenamefont
  {Ben~Dor}, \citenamefont {Yochelis}, \citenamefont {Radko}, \citenamefont
  {Vankayala}, \citenamefont {Capua}, \citenamefont {Capua}, \citenamefont
  {Yang}, \citenamefont {Baczewski}, \citenamefont {Parkin}, \citenamefont
  {Naaman},\ and\ \citenamefont {Paltiel}}]{ben_dor_magnetization_2017}%
  \BibitemOpen
  \bibfield  {author} {\bibinfo {author} {\bibfnamefont {O.}~\bibnamefont
  {Ben~Dor}}, \bibinfo {author} {\bibfnamefont {S.}~\bibnamefont {Yochelis}},
  \bibinfo {author} {\bibfnamefont {A.}~\bibnamefont {Radko}}, \bibinfo
  {author} {\bibfnamefont {K.}~\bibnamefont {Vankayala}}, \bibinfo {author}
  {\bibfnamefont {E.}~\bibnamefont {Capua}}, \bibinfo {author} {\bibfnamefont
  {A.}~\bibnamefont {Capua}}, \bibinfo {author} {\bibfnamefont {S.-H.}\
  \bibnamefont {Yang}}, \bibinfo {author} {\bibfnamefont {L.~T.}\ \bibnamefont
  {Baczewski}}, \bibinfo {author} {\bibfnamefont {S.~S.~P.}\ \bibnamefont
  {Parkin}}, \bibinfo {author} {\bibfnamefont {R.}~\bibnamefont {Naaman}},\
  and\ \bibinfo {author} {\bibfnamefont {Y.}~\bibnamefont {Paltiel}},\ }\href
  {https://doi.org/10.1038/ncomms14567} {\bibfield  {journal} {\bibinfo
  {journal} {Nature Communications}\ }\textbf {\bibinfo {volume} {8}},\
  \bibinfo {pages} {14567} (\bibinfo {year} {2017})}\BibitemShut {NoStop}%
\bibitem [{\citenamefont {Naaman}\ \emph {et~al.}(2020)\citenamefont {Naaman},
  \citenamefont {Paltiel},\ and\ \citenamefont {Waldeck}}]{naaman_chiral_2020}%
  \BibitemOpen
  \bibfield  {author} {\bibinfo {author} {\bibfnamefont {R.}~\bibnamefont
  {Naaman}}, \bibinfo {author} {\bibfnamefont {Y.}~\bibnamefont {Paltiel}},\
  and\ \bibinfo {author} {\bibfnamefont {D.~H.}\ \bibnamefont {Waldeck}},\
  }\href {https://doi.org/10.1021/acs.accounts.0c00485} {\bibfield  {journal}
  {\bibinfo  {journal} {Accounts of Chemical Research}\ }\textbf {\bibinfo
  {volume} {53}},\ \bibinfo {pages} {2659} (\bibinfo {year}
  {2020})}\BibitemShut {NoStop}%
\bibitem [{\citenamefont {Abendroth}\ \emph {et~al.}(2019)\citenamefont
  {Abendroth}, \citenamefont {Stemer}, \citenamefont {Bloom}, \citenamefont
  {Roy}, \citenamefont {Naaman}, \citenamefont {Waldeck}, \citenamefont
  {Weiss},\ and\ \citenamefont {Mondal}}]{abendroth_spin_2019}%
  \BibitemOpen
  \bibfield  {author} {\bibinfo {author} {\bibfnamefont {J.~M.}\ \bibnamefont
  {Abendroth}}, \bibinfo {author} {\bibfnamefont {D.~M.}\ \bibnamefont
  {Stemer}}, \bibinfo {author} {\bibfnamefont {B.~P.}\ \bibnamefont {Bloom}},
  \bibinfo {author} {\bibfnamefont {P.}~\bibnamefont {Roy}}, \bibinfo {author}
  {\bibfnamefont {R.}~\bibnamefont {Naaman}}, \bibinfo {author} {\bibfnamefont
  {D.~H.}\ \bibnamefont {Waldeck}}, \bibinfo {author} {\bibfnamefont {P.~S.}\
  \bibnamefont {Weiss}},\ and\ \bibinfo {author} {\bibfnamefont {P.~C.}\
  \bibnamefont {Mondal}},\ }\href {https://doi.org/10.1021/acsnano.9b01876}
  {\bibfield  {journal} {\bibinfo  {journal} {ACS Nano}\ }\textbf {\bibinfo
  {volume} {13}},\ \bibinfo {pages} {4928} (\bibinfo {year}
  {2019})}\BibitemShut {NoStop}%
\bibitem [{\citenamefont {Carmeli}\ \emph {et~al.}(2014)\citenamefont
  {Carmeli}, \citenamefont {Kumar}, \citenamefont {Heifler}, \citenamefont
  {Carmeli},\ and\ \citenamefont {Naaman}}]{carmeli_spin_2014}%
  \BibitemOpen
  \bibfield  {author} {\bibinfo {author} {\bibfnamefont {I.}~\bibnamefont
  {Carmeli}}, \bibinfo {author} {\bibfnamefont {K.~S.}\ \bibnamefont {Kumar}},
  \bibinfo {author} {\bibfnamefont {O.}~\bibnamefont {Heifler}}, \bibinfo
  {author} {\bibfnamefont {C.}~\bibnamefont {Carmeli}},\ and\ \bibinfo {author}
  {\bibfnamefont {R.}~\bibnamefont {Naaman}},\ }\href
  {https://doi.org/10.1002/anie.201404382} {\bibfield  {journal} {\bibinfo
  {journal} {Angewandte Chemie International Edition}\ }\textbf {\bibinfo
  {volume} {53}},\ \bibinfo {pages} {8953} (\bibinfo {year}
  {2014})}\BibitemShut {NoStop}%
\bibitem [{\citenamefont {Michaeli}\ \emph {et~al.}(2016)\citenamefont
  {Michaeli}, \citenamefont {Kantor-Uriel}, \citenamefont {Naaman},\ and\
  \citenamefont {Waldeck}}]{michaeli_electrons_2016}%
  \BibitemOpen
  \bibfield  {author} {\bibinfo {author} {\bibfnamefont {K.}~\bibnamefont
  {Michaeli}}, \bibinfo {author} {\bibfnamefont {N.}~\bibnamefont
  {Kantor-Uriel}}, \bibinfo {author} {\bibfnamefont {R.}~\bibnamefont
  {Naaman}},\ and\ \bibinfo {author} {\bibfnamefont {D.~H.}\ \bibnamefont
  {Waldeck}},\ }\href {https://doi.org/10.1039/C6CS00369A} {\bibfield
  {journal} {\bibinfo  {journal} {Chemical Society Reviews}\ }\textbf {\bibinfo
  {volume} {45}},\ \bibinfo {pages} {6478} (\bibinfo {year}
  {2016})}\BibitemShut {NoStop}%
\bibitem [{\citenamefont {Naaman}\ \emph {et~al.}(2019)\citenamefont {Naaman},
  \citenamefont {Waldeck},\ and\ \citenamefont {Paltiel}}]{naaman_chiral_2019}%
  \BibitemOpen
  \bibfield  {author} {\bibinfo {author} {\bibfnamefont {R.}~\bibnamefont
  {Naaman}}, \bibinfo {author} {\bibfnamefont {D.~H.}\ \bibnamefont
  {Waldeck}},\ and\ \bibinfo {author} {\bibfnamefont {Y.}~\bibnamefont
  {Paltiel}},\ }\href {https://doi.org/10.1063/1.5125034} {\bibfield  {journal}
  {\bibinfo  {journal} {Applied Physics Letters}\ }\textbf {\bibinfo {volume}
  {115}},\ \bibinfo {pages} {133701} (\bibinfo {year} {2019})}\BibitemShut
  {NoStop}%
\bibitem [{\citenamefont {Ozturk}\ and\ \citenamefont
  {Sasselov}(2022)}]{ozturk_origins_2022}%
  \BibitemOpen
  \bibfield  {author} {\bibinfo {author} {\bibfnamefont {S.~F.}\ \bibnamefont
  {Ozturk}}\ and\ \bibinfo {author} {\bibfnamefont {D.~D.}\ \bibnamefont
  {Sasselov}},\ }\href {https://doi.org/10.1073/pnas.2204765119} {\bibfield
  {journal} {\bibinfo  {journal} {Proceedings of the National Academy of
  Sciences}\ }\textbf {\bibinfo {volume} {119}},\ \bibinfo {pages}
  {e2204765119} (\bibinfo {year} {2022})}\BibitemShut {NoStop}%
\bibitem [{\citenamefont
  {Lehmberg}(1970{\natexlab{a}})}]{lehmberg_radiation_1970}%
  \BibitemOpen
  \bibfield  {author} {\bibinfo {author} {\bibfnamefont {R.~H.}\ \bibnamefont
  {Lehmberg}},\ }\href {https://doi.org/10.1103/PhysRevA.2.883} {\bibfield
  {journal} {\bibinfo  {journal} {Physical Review A}\ }\textbf {\bibinfo
  {volume} {2}},\ \bibinfo {pages} {883} (\bibinfo {year}
  {1970}{\natexlab{a}})}\BibitemShut {NoStop}%
\bibitem [{\citenamefont
  {Lehmberg}(1970{\natexlab{b}})}]{lehmberg_radiation_1970_2}%
  \BibitemOpen
  \bibfield  {author} {\bibinfo {author} {\bibfnamefont {R.~H.}\ \bibnamefont
  {Lehmberg}},\ }\href {https://doi.org/10.1103/PhysRevA.2.889} {\bibfield
  {journal} {\bibinfo  {journal} {Physical Review A}\ }\textbf {\bibinfo
  {volume} {2}},\ \bibinfo {pages} {889} (\bibinfo {year}
  {1970}{\natexlab{b}})}\BibitemShut {NoStop}%
\bibitem [{\citenamefont {Olmos}\ \emph {et~al.}(2013)\citenamefont {Olmos},
  \citenamefont {Yu}, \citenamefont {Singh}, \citenamefont {Schreck},
  \citenamefont {Bongs},\ and\ \citenamefont
  {Lesanovsky}}]{olmos_long-range_2013}%
  \BibitemOpen
  \bibfield  {author} {\bibinfo {author} {\bibfnamefont {B.}~\bibnamefont
  {Olmos}}, \bibinfo {author} {\bibfnamefont {D.}~\bibnamefont {Yu}}, \bibinfo
  {author} {\bibfnamefont {Y.}~\bibnamefont {Singh}}, \bibinfo {author}
  {\bibfnamefont {F.}~\bibnamefont {Schreck}}, \bibinfo {author} {\bibfnamefont
  {K.}~\bibnamefont {Bongs}},\ and\ \bibinfo {author} {\bibfnamefont
  {I.}~\bibnamefont {Lesanovsky}},\ }\href
  {https://doi.org/10.1103/PhysRevLett.110.143602} {\bibfield  {journal}
  {\bibinfo  {journal} {Physical Review Letters}\ }\textbf {\bibinfo {volume}
  {110}},\ \bibinfo {pages} {143602} (\bibinfo {year} {2013})}\BibitemShut
  {NoStop}%
\bibitem [{\citenamefont {Gutiérrez-Jáuregui}\ and\ \citenamefont
  {Asenjo-Garcia}(2022)}]{gutierrez-jauregui_directional_2022}%
  \BibitemOpen
  \bibfield  {author} {\bibinfo {author} {\bibfnamefont {R.}~\bibnamefont
  {Gutiérrez-Jáuregui}}\ and\ \bibinfo {author} {\bibfnamefont
  {A.}~\bibnamefont {Asenjo-Garcia}},\ }\href
  {https://doi.org/10.1103/PhysRevA.105.043703} {\bibfield  {journal} {\bibinfo
   {journal} {Physical Review A}\ }\textbf {\bibinfo {volume} {105}},\ \bibinfo
  {pages} {043703} (\bibinfo {year} {2022})}\BibitemShut {NoStop}%
\bibitem [{\citenamefont {Jang}\ and\ \citenamefont
  {Mennucci}(2018)}]{jang_delocalized_2018}%
  \BibitemOpen
  \bibfield  {author} {\bibinfo {author} {\bibfnamefont {S.~J.}\ \bibnamefont
  {Jang}}\ and\ \bibinfo {author} {\bibfnamefont {B.}~\bibnamefont
  {Mennucci}},\ }\href {https://doi.org/10.1103/RevModPhys.90.035003}
  {\bibfield  {journal} {\bibinfo  {journal} {Reviews of Modern Physics}\
  }\textbf {\bibinfo {volume} {90}},\ \bibinfo {pages} {035003} (\bibinfo
  {year} {2018})}\BibitemShut {NoStop}%
\bibitem [{\citenamefont {Asenjo-Garcia}\ \emph {et~al.}(2017)\citenamefont
  {Asenjo-Garcia}, \citenamefont {Moreno-Cardoner}, \citenamefont {Albrecht},
  \citenamefont {Kimble},\ and\ \citenamefont
  {Chang}}]{asenjo-garcia_exponential_2017}%
  \BibitemOpen
  \bibfield  {author} {\bibinfo {author} {\bibfnamefont {A.}~\bibnamefont
  {Asenjo-Garcia}}, \bibinfo {author} {\bibfnamefont {M.}~\bibnamefont
  {Moreno-Cardoner}}, \bibinfo {author} {\bibfnamefont {A.}~\bibnamefont
  {Albrecht}}, \bibinfo {author} {\bibfnamefont {H.}~\bibnamefont {Kimble}},\
  and\ \bibinfo {author} {\bibfnamefont {D.}~\bibnamefont {Chang}},\ }\href
  {https://doi.org/10.1103/PhysRevX.7.031024} {\bibfield  {journal} {\bibinfo
  {journal} {Physical Review X}\ }\textbf {\bibinfo {volume} {7}},\ \bibinfo
  {pages} {031024} (\bibinfo {year} {2017})}\BibitemShut {NoStop}%
\bibitem [{\citenamefont {Reitz}\ \emph {et~al.}(2022)\citenamefont {Reitz},
  \citenamefont {Sommer},\ and\ \citenamefont
  {Genes}}]{reitz_cooperative_2022}%
  \BibitemOpen
  \bibfield  {author} {\bibinfo {author} {\bibfnamefont {M.}~\bibnamefont
  {Reitz}}, \bibinfo {author} {\bibfnamefont {C.}~\bibnamefont {Sommer}},\ and\
  \bibinfo {author} {\bibfnamefont {C.}~\bibnamefont {Genes}},\ }\href
  {https://doi.org/10.1103/PRXQuantum.3.010201} {\bibfield  {journal} {\bibinfo
   {journal} {PRX Quantum}\ }\textbf {\bibinfo {volume} {3}},\ \bibinfo {pages}
  {010201} (\bibinfo {year} {2022})}\BibitemShut {NoStop}%
\bibitem [{\citenamefont {MacDonald}\ \emph {et~al.}(2001)\citenamefont
  {MacDonald}, \citenamefont {Herbert}, \citenamefont {Zhang}, \citenamefont
  {Polgruto},\ and\ \citenamefont {Lu}}]{macdonald_solution_2001}%
  \BibitemOpen
  \bibfield  {author} {\bibinfo {author} {\bibfnamefont {D.}~\bibnamefont
  {MacDonald}}, \bibinfo {author} {\bibfnamefont {K.}~\bibnamefont {Herbert}},
  \bibinfo {author} {\bibfnamefont {X.}~\bibnamefont {Zhang}}, \bibinfo
  {author} {\bibfnamefont {T.}~\bibnamefont {Polgruto}},\ and\ \bibinfo
  {author} {\bibfnamefont {P.}~\bibnamefont {Lu}},\ }\href
  {https://doi.org/10.1006/jmbi.2001.4447} {\bibfield  {journal} {\bibinfo
  {journal} {Journal of Molecular Biology}\ }\textbf {\bibinfo {volume}
  {306}},\ \bibinfo {pages} {1081} (\bibinfo {year} {2001})}\BibitemShut
  {NoStop}%
\bibitem [{\citenamefont {Wilczek}\ and\ \citenamefont
  {Zee}(1984)}]{wilczek_appearance_1984}%
  \BibitemOpen
  \bibfield  {author} {\bibinfo {author} {\bibfnamefont {F.}~\bibnamefont
  {Wilczek}}\ and\ \bibinfo {author} {\bibfnamefont {A.}~\bibnamefont {Zee}},\
  }\href {https://doi.org/10.1103/PhysRevLett.52.2111} {\bibfield  {journal}
  {\bibinfo  {journal} {Physical Review Letters}\ }\textbf {\bibinfo {volume}
  {52}},\ \bibinfo {pages} {2111} (\bibinfo {year} {1984})}\BibitemShut
  {NoStop}%
\bibitem [{\citenamefont {Hafezi}\ \emph {et~al.}(2011)\citenamefont {Hafezi},
  \citenamefont {Demler}, \citenamefont {Lukin},\ and\ \citenamefont
  {Taylor}}]{hafezi_robust_2011}%
  \BibitemOpen
  \bibfield  {author} {\bibinfo {author} {\bibfnamefont {M.}~\bibnamefont
  {Hafezi}}, \bibinfo {author} {\bibfnamefont {E.~A.}\ \bibnamefont {Demler}},
  \bibinfo {author} {\bibfnamefont {M.~D.}\ \bibnamefont {Lukin}},\ and\
  \bibinfo {author} {\bibfnamefont {J.~M.}\ \bibnamefont {Taylor}},\ }\href
  {https://doi.org/10.1038/nphys2063} {\bibfield  {journal} {\bibinfo
  {journal} {Nature Physics}\ }\textbf {\bibinfo {volume} {7}},\ \bibinfo
  {pages} {907} (\bibinfo {year} {2011})}\BibitemShut {NoStop}%
\bibitem [{\citenamefont {Syzranov}\ \emph {et~al.}(2014)\citenamefont
  {Syzranov}, \citenamefont {Wall}, \citenamefont {Gurarie},\ and\
  \citenamefont {Rey}}]{syzranov_spinorbital_2014}%
  \BibitemOpen
  \bibfield  {author} {\bibinfo {author} {\bibfnamefont {S.~V.}\ \bibnamefont
  {Syzranov}}, \bibinfo {author} {\bibfnamefont {M.~L.}\ \bibnamefont {Wall}},
  \bibinfo {author} {\bibfnamefont {V.}~\bibnamefont {Gurarie}},\ and\ \bibinfo
  {author} {\bibfnamefont {A.~M.}\ \bibnamefont {Rey}},\ }\href
  {https://doi.org/10.1038/ncomms6391} {\bibfield  {journal} {\bibinfo
  {journal} {Nature Communications}\ }\textbf {\bibinfo {volume} {5}},\
  \bibinfo {pages} {5391} (\bibinfo {year} {2014})}\BibitemShut {NoStop}%
\bibitem [{\citenamefont {May}\ and\ \citenamefont
  {Kühn}(2011)}]{may_charge_2011}%
  \BibitemOpen
  \bibfield  {author} {\bibinfo {author} {\bibfnamefont {V.}~\bibnamefont
  {May}}\ and\ \bibinfo {author} {\bibfnamefont {O.}~\bibnamefont {Kühn}},\
  }\href {https://doi.org/10.1002/9783527633791} {\emph {\bibinfo {title}
  {Charge and {Energy} {Transfer} {Dynamics} in {Molecular} {Systems}}}},\
  \bibinfo {edition} {1st}\ ed.\ (\bibinfo  {publisher} {Wiley},\ \bibinfo
  {year} {2011})\BibitemShut {NoStop}%
\bibitem [{\citenamefont {Peter}\ \emph {et~al.}(tted)\citenamefont {Peter},
  \citenamefont {Ostermann},\ and\ \citenamefont {Yelin}}]{peter_PRA}%
  \BibitemOpen
  \bibfield  {author} {\bibinfo {author} {\bibfnamefont {J.~S.}\ \bibnamefont
  {Peter}}, \bibinfo {author} {\bibfnamefont {S.}~\bibnamefont {Ostermann}},\
  and\ \bibinfo {author} {\bibfnamefont {S.~F.}\ \bibnamefont {Yelin}},\
  }\href@noop {} {\bibfield  {journal} {\bibinfo  {journal} {Phys. Rev. A}\ }
  (\bibinfo {year} {submitted})}\BibitemShut {NoStop}%
\bibitem [{\citenamefont {Kane}(2013)}]{kane_topological_2013}%
  \BibitemOpen
  \bibfield  {author} {\bibinfo {author} {\bibfnamefont {C.}~\bibnamefont
  {Kane}},\ }in\ \href {https://doi.org/10.1016/B978-0-444-63314-9.00001-9}
  {\emph {\bibinfo {booktitle} {Contemporary {Concepts} of {Condensed} {Matter}
  {Science}}}},\ Vol.~\bibinfo {volume} {6}\ (\bibinfo  {publisher}
  {Elsevier},\ \bibinfo {year} {2013})\ pp.\ \bibinfo {pages}
  {3--34}\BibitemShut {NoStop}%
\bibitem [{\citenamefont {Halperin}(1982)}]{halperin_quantized_1982}%
  \BibitemOpen
  \bibfield  {author} {\bibinfo {author} {\bibfnamefont {B.~I.}\ \bibnamefont
  {Halperin}},\ }\href {https://doi.org/10.1103/PhysRevB.25.2185} {\bibfield
  {journal} {\bibinfo  {journal} {Physical Review B}\ }\textbf {\bibinfo
  {volume} {25}},\ \bibinfo {pages} {2185} (\bibinfo {year}
  {1982})}\BibitemShut {NoStop}%
\bibitem [{\citenamefont {Atala}\ \emph {et~al.}(2013)\citenamefont {Atala},
  \citenamefont {Aidelsburger}, \citenamefont {Barreiro}, \citenamefont
  {Abanin}, \citenamefont {Kitagawa}, \citenamefont {Demler},\ and\
  \citenamefont {Bloch}}]{atala_direct_2013}%
  \BibitemOpen
  \bibfield  {author} {\bibinfo {author} {\bibfnamefont {M.}~\bibnamefont
  {Atala}}, \bibinfo {author} {\bibfnamefont {M.}~\bibnamefont {Aidelsburger}},
  \bibinfo {author} {\bibfnamefont {J.~T.}\ \bibnamefont {Barreiro}}, \bibinfo
  {author} {\bibfnamefont {D.}~\bibnamefont {Abanin}}, \bibinfo {author}
  {\bibfnamefont {T.}~\bibnamefont {Kitagawa}}, \bibinfo {author}
  {\bibfnamefont {E.}~\bibnamefont {Demler}},\ and\ \bibinfo {author}
  {\bibfnamefont {I.}~\bibnamefont {Bloch}},\ }\href
  {https://doi.org/10.1038/nphys2790} {\bibfield  {journal} {\bibinfo
  {journal} {Nature Physics}\ }\textbf {\bibinfo {volume} {9}},\ \bibinfo
  {pages} {795} (\bibinfo {year} {2013})}\BibitemShut {NoStop}%
\bibitem [{\citenamefont {Zak}(1989)}]{zak_berrys_1989}%
  \BibitemOpen
  \bibfield  {author} {\bibinfo {author} {\bibfnamefont {J.}~\bibnamefont
  {Zak}},\ }\href {https://doi.org/10.1103/PhysRevLett.62.2747} {\bibfield
  {journal} {\bibinfo  {journal} {Physical Review Letters}\ }\textbf {\bibinfo
  {volume} {62}},\ \bibinfo {pages} {2747} (\bibinfo {year}
  {1989})}\BibitemShut {NoStop}%
\bibitem [{\citenamefont {Liu}\ \emph {et~al.}(2021)\citenamefont {Liu},
  \citenamefont {Xiao}, \citenamefont {Koo},\ and\ \citenamefont
  {Yan}}]{liu_chirality-driven_2021}%
  \BibitemOpen
  \bibfield  {author} {\bibinfo {author} {\bibfnamefont {Y.}~\bibnamefont
  {Liu}}, \bibinfo {author} {\bibfnamefont {J.}~\bibnamefont {Xiao}}, \bibinfo
  {author} {\bibfnamefont {J.}~\bibnamefont {Koo}},\ and\ \bibinfo {author}
  {\bibfnamefont {B.}~\bibnamefont {Yan}},\ }\href
  {https://doi.org/10.1038/s41563-021-00924-5} {\bibfield  {journal} {\bibinfo
  {journal} {Nature Materials}\ }\textbf {\bibinfo {volume} {20}},\ \bibinfo
  {pages} {638} (\bibinfo {year} {2021})}\BibitemShut {NoStop}%
\bibitem [{\citenamefont {Ayuso}\ \emph {et~al.}(2022)\citenamefont {Ayuso},
  \citenamefont {Ordonez},\ and\ \citenamefont
  {Smirnova}}]{ayuso_ultrafast_2022}%
  \BibitemOpen
  \bibfield  {author} {\bibinfo {author} {\bibfnamefont {D.}~\bibnamefont
  {Ayuso}}, \bibinfo {author} {\bibfnamefont {A.~F.}\ \bibnamefont {Ordonez}},\
  and\ \bibinfo {author} {\bibfnamefont {O.}~\bibnamefont {Smirnova}},\ }\href
  {https://doi.org/10.1039/D2CP01009G} {\bibfield  {journal} {\bibinfo
  {journal} {Physical Chemistry Chemical Physics}\ }\textbf {\bibinfo {volume}
  {24}},\ \bibinfo {pages} {26962} (\bibinfo {year} {2022})}\BibitemShut
  {NoStop}%
\bibitem [{\citenamefont {Ordonez}\ and\ \citenamefont
  {Smirnova}(2018)}]{ordonez_generalized_2018}%
  \BibitemOpen
  \bibfield  {author} {\bibinfo {author} {\bibfnamefont {A.~F.}\ \bibnamefont
  {Ordonez}}\ and\ \bibinfo {author} {\bibfnamefont {O.}~\bibnamefont
  {Smirnova}},\ }\href {https://doi.org/10.1103/PhysRevA.98.063428} {\bibfield
  {journal} {\bibinfo  {journal} {Physical Review A}\ }\textbf {\bibinfo
  {volume} {98}},\ \bibinfo {pages} {063428} (\bibinfo {year}
  {2018})}\BibitemShut {NoStop}%
\bibitem [{\citenamefont {Nielsen}\ \emph {et~al.}(2012)\citenamefont
  {Nielsen}, \citenamefont {Hoffmann},\ and\ \citenamefont
  {Nielsen}}]{nielsen_probing_2012}%
  \BibitemOpen
  \bibfield  {author} {\bibinfo {author} {\bibfnamefont {L.~M.}\ \bibnamefont
  {Nielsen}}, \bibinfo {author} {\bibfnamefont {S.~V.}\ \bibnamefont
  {Hoffmann}},\ and\ \bibinfo {author} {\bibfnamefont {S.~B.}\ \bibnamefont
  {Nielsen}},\ }\href {https://doi.org/10.1039/c2cc35201j} {\bibfield
  {journal} {\bibinfo  {journal} {Chemical Communications}\ }\textbf {\bibinfo
  {volume} {48}},\ \bibinfo {pages} {10425} (\bibinfo {year}
  {2012})}\BibitemShut {NoStop}%
\bibitem [{\citenamefont {Nielsen}\ \emph {et~al.}(2013)\citenamefont
  {Nielsen}, \citenamefont {Hoffmann},\ and\ \citenamefont
  {Brøndsted~Nielsen}}]{nielsen_electronic_2013}%
  \BibitemOpen
  \bibfield  {author} {\bibinfo {author} {\bibfnamefont {L.~M.}\ \bibnamefont
  {Nielsen}}, \bibinfo {author} {\bibfnamefont {S.~V.}\ \bibnamefont
  {Hoffmann}},\ and\ \bibinfo {author} {\bibfnamefont {S.}~\bibnamefont
  {Brøndsted~Nielsen}},\ }\href {https://doi.org/10.1039/c3pp25438k}
  {\bibfield  {journal} {\bibinfo  {journal} {Photochemical \& Photobiological
  Sciences}\ }\textbf {\bibinfo {volume} {12}},\ \bibinfo {pages} {1273}
  (\bibinfo {year} {2013})}\BibitemShut {NoStop}%
\bibitem [{\citenamefont {Buchvarov}\ \emph {et~al.}(2007)\citenamefont
  {Buchvarov}, \citenamefont {Wang}, \citenamefont {Raytchev}, \citenamefont
  {Trifonov},\ and\ \citenamefont {Fiebig}}]{buchvarov_electronic_2007}%
  \BibitemOpen
  \bibfield  {author} {\bibinfo {author} {\bibfnamefont {I.}~\bibnamefont
  {Buchvarov}}, \bibinfo {author} {\bibfnamefont {Q.}~\bibnamefont {Wang}},
  \bibinfo {author} {\bibfnamefont {M.}~\bibnamefont {Raytchev}}, \bibinfo
  {author} {\bibfnamefont {A.}~\bibnamefont {Trifonov}},\ and\ \bibinfo
  {author} {\bibfnamefont {T.}~\bibnamefont {Fiebig}},\ }\href
  {https://doi.org/10.1073/pnas.0606757104} {\bibfield  {journal} {\bibinfo
  {journal} {Proceedings of the National Academy of Sciences}\ }\textbf
  {\bibinfo {volume} {104}},\ \bibinfo {pages} {4794} (\bibinfo {year}
  {2007})}\BibitemShut {NoStop}%
\bibitem [{\citenamefont {Blackmond}(2010)}]{blackmond_origin_2010}%
  \BibitemOpen
  \bibfield  {author} {\bibinfo {author} {\bibfnamefont {D.~G.}\ \bibnamefont
  {Blackmond}},\ }\href {https://doi.org/10.1101/cshperspect.a002147}
  {\bibfield  {journal} {\bibinfo  {journal} {Cold Spring Harbor Perspectives
  in Biology}\ }\textbf {\bibinfo {volume} {2}},\ \bibinfo {pages} {a002147}
  (\bibinfo {year} {2010})}\BibitemShut {NoStop}%
\bibitem [{\citenamefont {Hein}\ and\ \citenamefont
  {Blackmond}(2012)}]{hein_origin_2012}%
  \BibitemOpen
  \bibfield  {author} {\bibinfo {author} {\bibfnamefont {J.~E.}\ \bibnamefont
  {Hein}}\ and\ \bibinfo {author} {\bibfnamefont {D.~G.}\ \bibnamefont
  {Blackmond}},\ }\href {https://doi.org/10.1021/ar200316n} {\bibfield
  {journal} {\bibinfo  {journal} {Accounts of Chemical Research}\ }\textbf
  {\bibinfo {volume} {45}},\ \bibinfo {pages} {2045} (\bibinfo {year}
  {2012})}\BibitemShut {NoStop}%
\bibitem [{\citenamefont {Bailey}\ \emph {et~al.}(1998)\citenamefont {Bailey},
  \citenamefont {Chrysostomou}, \citenamefont {Hough}, \citenamefont
  {Gledhill}, \citenamefont {McCall}, \citenamefont {Clark}, \citenamefont
  {Ménard},\ and\ \citenamefont {Tamura}}]{bailey_circular_1998}%
  \BibitemOpen
  \bibfield  {author} {\bibinfo {author} {\bibfnamefont {J.}~\bibnamefont
  {Bailey}}, \bibinfo {author} {\bibfnamefont {A.}~\bibnamefont
  {Chrysostomou}}, \bibinfo {author} {\bibfnamefont {J.~H.}\ \bibnamefont
  {Hough}}, \bibinfo {author} {\bibfnamefont {T.~M.}\ \bibnamefont {Gledhill}},
  \bibinfo {author} {\bibfnamefont {A.}~\bibnamefont {McCall}}, \bibinfo
  {author} {\bibfnamefont {S.}~\bibnamefont {Clark}}, \bibinfo {author}
  {\bibfnamefont {F.}~\bibnamefont {Ménard}},\ and\ \bibinfo {author}
  {\bibfnamefont {M.}~\bibnamefont {Tamura}},\ }\href
  {https://doi.org/10.1126/science.281.5377.672} {\bibfield  {journal}
  {\bibinfo  {journal} {Science}\ }\textbf {\bibinfo {volume} {281}},\ \bibinfo
  {pages} {672} (\bibinfo {year} {1998})}\BibitemShut {NoStop}%
\bibitem [{\citenamefont {Fukue}\ \emph {et~al.}(2010)\citenamefont {Fukue},
  \citenamefont {Tamura}, \citenamefont {Kandori}, \citenamefont {Kusakabe},
  \citenamefont {Hough}, \citenamefont {Bailey}, \citenamefont {Whittet},
  \citenamefont {Lucas}, \citenamefont {Nakajima},\ and\ \citenamefont
  {Hashimoto}}]{fukue_extended_2010}%
  \BibitemOpen
  \bibfield  {author} {\bibinfo {author} {\bibfnamefont {T.}~\bibnamefont
  {Fukue}}, \bibinfo {author} {\bibfnamefont {M.}~\bibnamefont {Tamura}},
  \bibinfo {author} {\bibfnamefont {R.}~\bibnamefont {Kandori}}, \bibinfo
  {author} {\bibfnamefont {N.}~\bibnamefont {Kusakabe}}, \bibinfo {author}
  {\bibfnamefont {J.~H.}\ \bibnamefont {Hough}}, \bibinfo {author}
  {\bibfnamefont {J.}~\bibnamefont {Bailey}}, \bibinfo {author} {\bibfnamefont
  {D.~C.~B.}\ \bibnamefont {Whittet}}, \bibinfo {author} {\bibfnamefont
  {P.~W.}\ \bibnamefont {Lucas}}, \bibinfo {author} {\bibfnamefont
  {Y.}~\bibnamefont {Nakajima}},\ and\ \bibinfo {author} {\bibfnamefont
  {J.}~\bibnamefont {Hashimoto}},\ }\href
  {https://doi.org/10.1007/s11084-010-9206-1} {\bibfield  {journal} {\bibinfo
  {journal} {Origins of Life and Evolution of the Biosphere}\ }\textbf
  {\bibinfo {volume} {40}},\ \bibinfo {pages} {335} (\bibinfo {year}
  {2010})}\BibitemShut {NoStop}%
\bibitem [{\citenamefont {Kwon}\ \emph {et~al.}(2013)\citenamefont {Kwon},
  \citenamefont {Tamura}, \citenamefont {Lucas}, \citenamefont {Hashimoto},
  \citenamefont {Kusakabe}, \citenamefont {Kandori}, \citenamefont {Nakajima},
  \citenamefont {Nagayama}, \citenamefont {Nagata},\ and\ \citenamefont
  {Hough}}]{kwon_near-infrared_2013}%
  \BibitemOpen
  \bibfield  {author} {\bibinfo {author} {\bibfnamefont {J.}~\bibnamefont
  {Kwon}}, \bibinfo {author} {\bibfnamefont {M.}~\bibnamefont {Tamura}},
  \bibinfo {author} {\bibfnamefont {P.~W.}\ \bibnamefont {Lucas}}, \bibinfo
  {author} {\bibfnamefont {J.}~\bibnamefont {Hashimoto}}, \bibinfo {author}
  {\bibfnamefont {N.}~\bibnamefont {Kusakabe}}, \bibinfo {author}
  {\bibfnamefont {R.}~\bibnamefont {Kandori}}, \bibinfo {author} {\bibfnamefont
  {Y.}~\bibnamefont {Nakajima}}, \bibinfo {author} {\bibfnamefont
  {T.}~\bibnamefont {Nagayama}}, \bibinfo {author} {\bibfnamefont
  {T.}~\bibnamefont {Nagata}},\ and\ \bibinfo {author} {\bibfnamefont {J.~H.}\
  \bibnamefont {Hough}},\ }\href {https://doi.org/10.1088/2041-8205/765/1/L6}
  {\bibfield  {journal} {\bibinfo  {journal} {The Astrophysical Journal}\
  }\textbf {\bibinfo {volume} {765}},\ \bibinfo {pages} {L6} (\bibinfo {year}
  {2013})}\BibitemShut {NoStop}%
\bibitem [{\citenamefont {Hadidi}\ \emph {et~al.}(2018)\citenamefont {Hadidi},
  \citenamefont {Bozanic}, \citenamefont {Garcia},\ and\ \citenamefont
  {Nahon}}]{hadidi_electron_2018}%
  \BibitemOpen
  \bibfield  {author} {\bibinfo {author} {\bibfnamefont {R.}~\bibnamefont
  {Hadidi}}, \bibinfo {author} {\bibfnamefont {D.~K.}\ \bibnamefont {Bozanic}},
  \bibinfo {author} {\bibfnamefont {G.~A.}\ \bibnamefont {Garcia}},\ and\
  \bibinfo {author} {\bibfnamefont {L.}~\bibnamefont {Nahon}},\ }\href
  {https://doi.org/10.1080/23746149.2018.1477530} {\bibfield  {journal}
  {\bibinfo  {journal} {Advances in Physics: X}\ }\textbf {\bibinfo {volume}
  {3}},\ \bibinfo {pages} {1477530} (\bibinfo {year} {2018})}\BibitemShut
  {NoStop}%
\bibitem [{\citenamefont {Garcia}\ \emph {et~al.}(2019)\citenamefont {Garcia},
  \citenamefont {Meinert}, \citenamefont {Sugahara}, \citenamefont {Jones},
  \citenamefont {Hoffmann},\ and\ \citenamefont
  {Meierhenrich}}]{garcia_astrophysical_2019}%
  \BibitemOpen
  \bibfield  {author} {\bibinfo {author} {\bibfnamefont {A.~D.}\ \bibnamefont
  {Garcia}}, \bibinfo {author} {\bibfnamefont {C.}~\bibnamefont {Meinert}},
  \bibinfo {author} {\bibfnamefont {H.}~\bibnamefont {Sugahara}}, \bibinfo
  {author} {\bibfnamefont {N.~C.}\ \bibnamefont {Jones}}, \bibinfo {author}
  {\bibfnamefont {S.~V.}\ \bibnamefont {Hoffmann}},\ and\ \bibinfo {author}
  {\bibfnamefont {U.~J.}\ \bibnamefont {Meierhenrich}},\ }\href
  {https://doi.org/10.3390/life9010029} {\bibfield  {journal} {\bibinfo
  {journal} {Life}\ }\textbf {\bibinfo {volume} {9}},\ \bibinfo {pages} {29}
  (\bibinfo {year} {2019})}\BibitemShut {NoStop}%
\bibitem [{\citenamefont {Jorissen}\ and\ \citenamefont
  {Cerf}(2002)}]{jorissen_asymmetric_2002}%
  \BibitemOpen
  \bibfield  {author} {\bibinfo {author} {\bibfnamefont {A.}~\bibnamefont
  {Jorissen}}\ and\ \bibinfo {author} {\bibfnamefont {C.}~\bibnamefont
  {Cerf}},\ }\href {https://doi.org/https://doi.org/10.1023/A:1016087202273}
  {\bibfield  {journal} {\bibinfo  {journal} {Origins of Life and Evolution of
  the Biosphere}\ }\textbf {\bibinfo {volume} {32}},\ \bibinfo {pages} {129}
  (\bibinfo {year} {2002})}\BibitemShut {NoStop}%
\bibitem [{\citenamefont {Balavoine}\ \emph {et~al.}(1974)\citenamefont
  {Balavoine}, \citenamefont {Moradpour},\ and\ \citenamefont
  {Kagan}}]{balavoine_preparation_1974}%
  \BibitemOpen
  \bibfield  {author} {\bibinfo {author} {\bibfnamefont {G.}~\bibnamefont
  {Balavoine}}, \bibinfo {author} {\bibfnamefont {A.}~\bibnamefont
  {Moradpour}},\ and\ \bibinfo {author} {\bibfnamefont {H.~B.}\ \bibnamefont
  {Kagan}},\ }\href {https://doi.org/10.1021/ja00823a023} {\bibfield  {journal}
  {\bibinfo  {journal} {Journal of the American Chemical Society}\ }\textbf
  {\bibinfo {volume} {96}},\ \bibinfo {pages} {5152} (\bibinfo {year}
  {1974})}\BibitemShut {NoStop}%
\bibitem [{\citenamefont {Flores}\ \emph {et~al.}(1977)\citenamefont {Flores},
  \citenamefont {Bonner},\ and\ \citenamefont
  {Massey}}]{flores_asymmetric_1977}%
  \BibitemOpen
  \bibfield  {author} {\bibinfo {author} {\bibfnamefont {J.~J.}\ \bibnamefont
  {Flores}}, \bibinfo {author} {\bibfnamefont {W.~A.}\ \bibnamefont {Bonner}},\
  and\ \bibinfo {author} {\bibfnamefont {G.~A.}\ \bibnamefont {Massey}},\
  }\href {https://doi.org/10.1021/ja00453a018} {\bibfield  {journal} {\bibinfo
  {journal} {Journal of the American Chemical Society}\ }\textbf {\bibinfo
  {volume} {99}},\ \bibinfo {pages} {3622} (\bibinfo {year}
  {1977})}\BibitemShut {NoStop}%
\bibitem [{\citenamefont {Meierhenrich}\ \emph {et~al.}(2005)\citenamefont
  {Meierhenrich}, \citenamefont {Nahon}, \citenamefont {Alcaraz}, \citenamefont
  {Bredehöft}, \citenamefont {Hoffmann}, \citenamefont {Barbier},\ and\
  \citenamefont {Brack}}]{meierhenrich_asymmetric_2005}%
  \BibitemOpen
  \bibfield  {author} {\bibinfo {author} {\bibfnamefont {U.~J.}\ \bibnamefont
  {Meierhenrich}}, \bibinfo {author} {\bibfnamefont {L.}~\bibnamefont {Nahon}},
  \bibinfo {author} {\bibfnamefont {C.}~\bibnamefont {Alcaraz}}, \bibinfo
  {author} {\bibfnamefont {J.~H.}\ \bibnamefont {Bredehöft}}, \bibinfo
  {author} {\bibfnamefont {S.~V.}\ \bibnamefont {Hoffmann}}, \bibinfo {author}
  {\bibfnamefont {B.}~\bibnamefont {Barbier}},\ and\ \bibinfo {author}
  {\bibfnamefont {A.}~\bibnamefont {Brack}},\ }\href
  {https://doi.org/10.1002/anie.200501311} {\bibfield  {journal} {\bibinfo
  {journal} {Angewandte Chemie International Edition}\ }\textbf {\bibinfo
  {volume} {44}},\ \bibinfo {pages} {5630} (\bibinfo {year}
  {2005})}\BibitemShut {NoStop}%
\bibitem [{\citenamefont {Meinert}\ \emph {et~al.}(2014)\citenamefont
  {Meinert}, \citenamefont {Hoffmann}, \citenamefont {Cassam-Chenaï},
  \citenamefont {Evans}, \citenamefont {Giri}, \citenamefont {Nahon},\ and\
  \citenamefont {Meierhenrich}}]{meinert_photonenergy-controlled_2014}%
  \BibitemOpen
  \bibfield  {author} {\bibinfo {author} {\bibfnamefont {C.}~\bibnamefont
  {Meinert}}, \bibinfo {author} {\bibfnamefont {S.~V.}\ \bibnamefont
  {Hoffmann}}, \bibinfo {author} {\bibfnamefont {P.}~\bibnamefont
  {Cassam-Chenaï}}, \bibinfo {author} {\bibfnamefont {A.~C.}\ \bibnamefont
  {Evans}}, \bibinfo {author} {\bibfnamefont {C.}~\bibnamefont {Giri}},
  \bibinfo {author} {\bibfnamefont {L.}~\bibnamefont {Nahon}},\ and\ \bibinfo
  {author} {\bibfnamefont {U.~J.}\ \bibnamefont {Meierhenrich}},\ }\href
  {https://doi.org/10.1002/anie.201307855} {\bibfield  {journal} {\bibinfo
  {journal} {Angewandte Chemie International Edition}\ }\textbf {\bibinfo
  {volume} {53}},\ \bibinfo {pages} {210} (\bibinfo {year} {2014})}\BibitemShut
  {NoStop}%
\end{thebibliography}%

\end{document}

\title{Supplemental Material for ``Chirality Dependent Photon Transport and Helical Superradiance"}

\author{Jonah S. Peter}
% \email{jonahpeter@g.harvard.edu}
\affiliation{Department of Physics, Harvard University, Cambridge, Massachusetts 02138, USA}
\affiliation{Biophysics Program, Harvard University, Boston, Massachusetts 02115, USA}
\author{Stefan Ostermann}
\affiliation{Department of Physics, Harvard University, Cambridge, Massachusetts 02138, USA}
\author{Susanne F. Yelin}
\affiliation{Department of Physics, Harvard University, Cambridge, Massachusetts 02138, USA}

\maketitle

\begin{figure}[h]
\centering 
\renewcommand\figurename{FIG.}
\includegraphics[width=0.7\columnwidth]{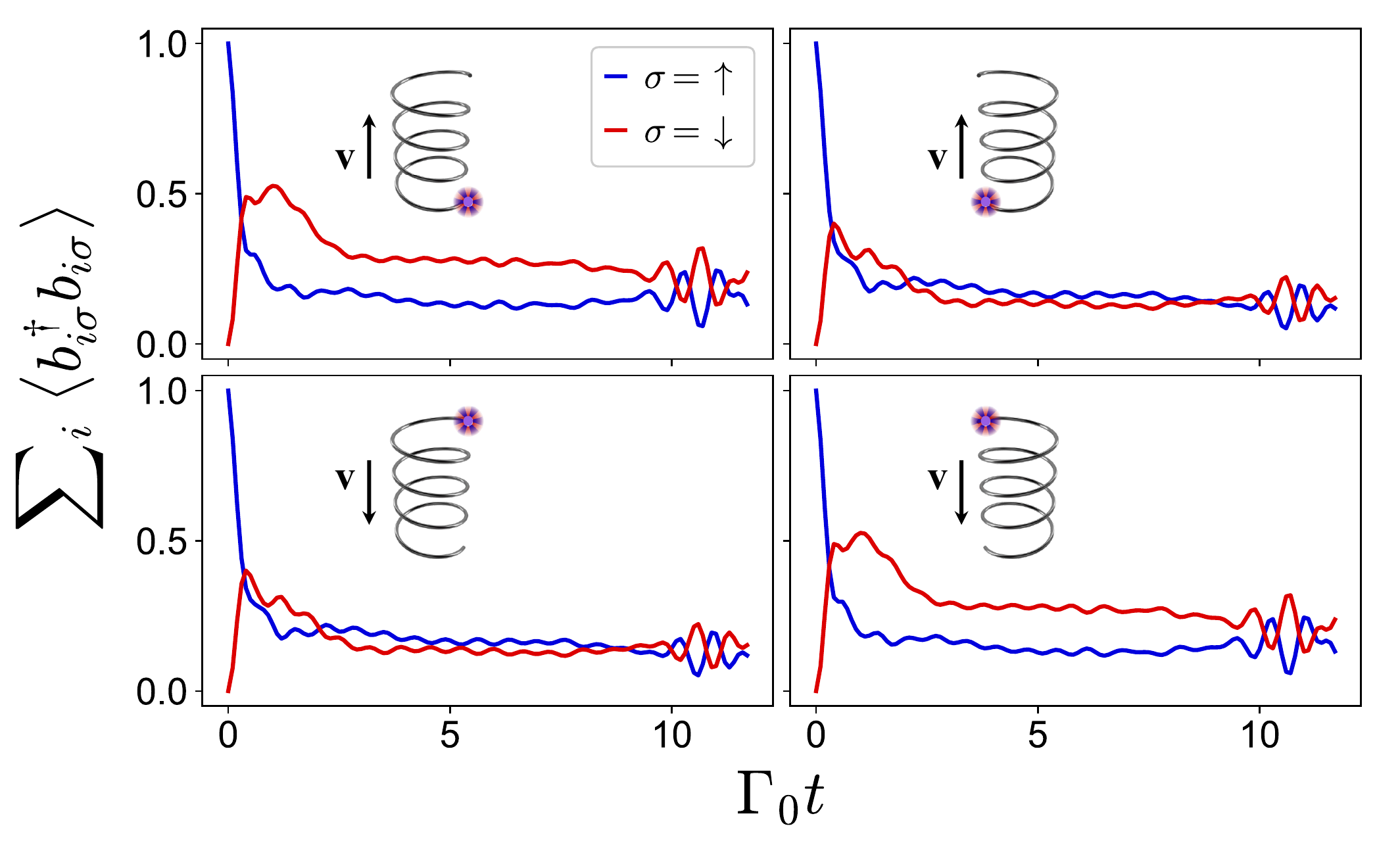}
\caption{Preferential excitation of one chirality from an unequally polarized initial state. Panels correspond to the same set-ups as in Fig. 2(a) of the main text but for initial states $\rho_0 = \ket{\uparrow_i}\bra{\uparrow_i}$ corresponding to a polarized spin up excitation localized at the bottom (top two panels) or top (bottom two panels) of the helix.}
\label{fig:polarized}
\end{figure}

\begin{figure}[h]
\centering 
\renewcommand\figurename{FIG.}
\includegraphics[width=0.7\columnwidth]{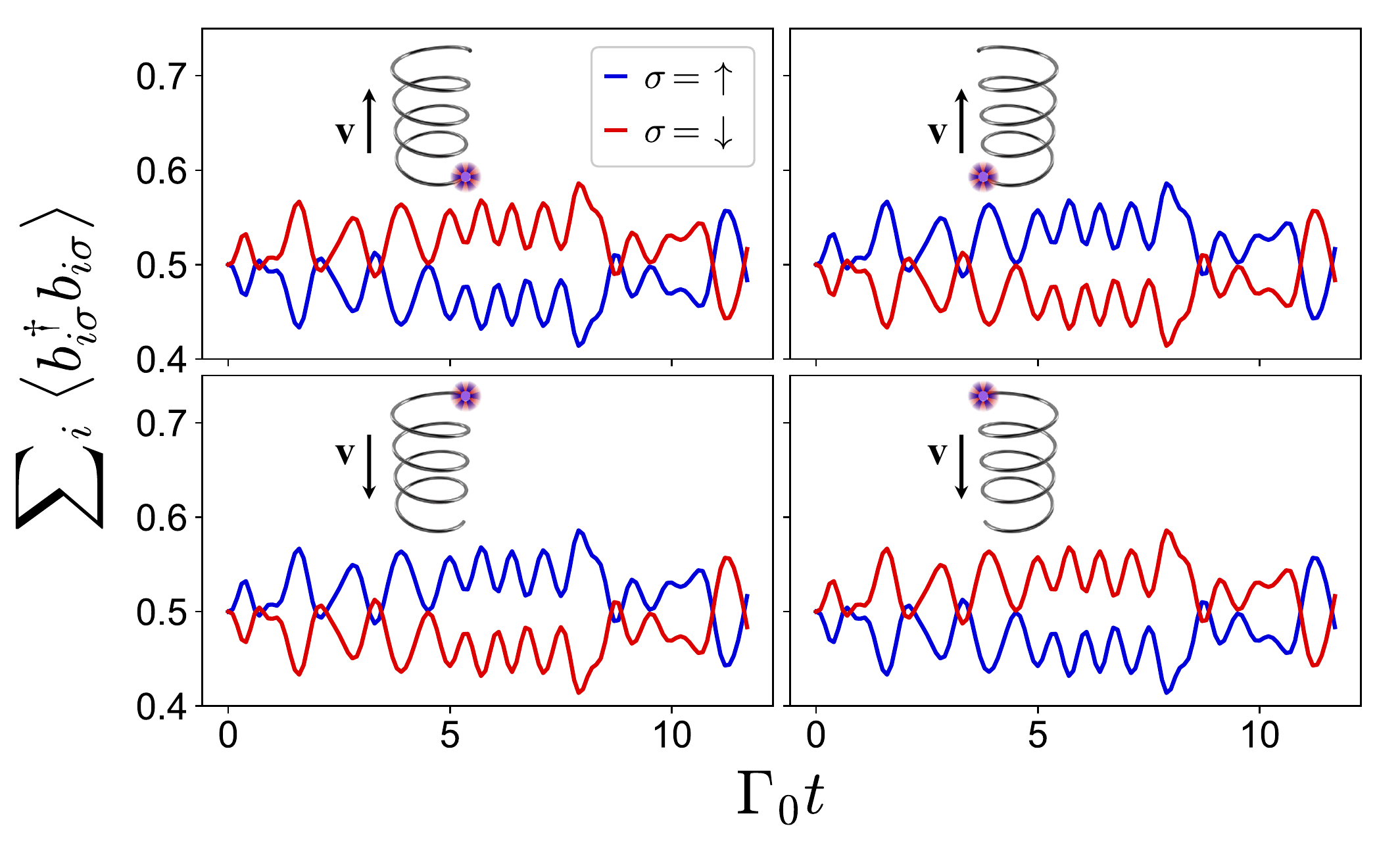}
\caption{Helicity dependent chiral transport without dissipation. Panels correspond to the same set-ups as in Fig. 2(a) of the main text but without the influence of the anti-Hermitian term in Eq. (2) [i.e., time evolution with only the Hamiltonian (1)]. SOC resulting from purely unitary dynamics is sufficient to achieve helicity dependent chiral transport.}
\label{fig:hermitian}
\end{figure}